\documentclass{sig-alternate}

\usepackage{color}
\usepackage{colortbl}
\usepackage{graphicx}
\usepackage{fancyvrb}
\usepackage{wasysym}
\usepackage{rotating}
\usepackage{multirow}
\usepackage{url}
\usepackage[lined,ruled,linesnumbered]{algorithm2e}
\usepackage{tabularx}
\usepackage{hyperref}

\hypersetup{
   pdfauthor={Martin Przyjaciel-Zablocki, Alexander Sch\"atzle, Thomas Hornung, Christopher Dorner, Georg Lausen},
   pdftitle={Cascading Map-Side Joins over HBase for Scalable Join Processing},
   pdfkeywords={MapReduce, SPARQL basic graph pattern, NoSQL, Map-Side Join, HBase, Semantic Web, MAPSIN},
   pdfsubject={One of the major challenges in large-scale data processing with MapReduce is the smart computation of joins. Since Semantic Web datasets published in RDF have increased rapidly over the last few years, scalable join techniques become an important issue for SPARQL query processing as well. In this paper, we introduce the Map-Side Index Nested Loop Join (MAPSIN join) which combines scalable indexing capabilities of NoSQL storage systems like HBase, that suffer from an insufficient distributed processing layer, with MapReduce, which in turn does not provide appropriate storage structures for efficient large-scale join processing. While retaining the flexibility of commonly used reduce-side joins, we leverage the effectiveness of map-side joins without any changes to the underlying framework. We demonstrate the significant benefits of MAPSIN joins for the processing of SPARQL basic graph patterns on large RDF datasets by an evaluation with the LUBM and SP2Bench benchmarks. For most queries, MAPSIN join based query execution outperforms reduce-side join based execution by an order of magnitude.}
}

\makeatletter
\let\@copyrightspace\relax
\makeatother

\begin{document}
\title{Cascading Map-Side Joins over HBase\\for Scalable Join Processing}

\numberofauthors{5} 
\author{
\alignauthor Martin Przyjaciel-Zablocki\\
       \affaddr{University of Freiburg, Germany}\\
       \email{zablocki@informatik.uni-freiburg.de}
\alignauthor Alexander Sch\"{a}tzle\\
       \affaddr{University of Freiburg, Germany}\\
       \email{schaetzle@informatik.uni-freiburg.de}
\alignauthor Thomas Hornung\\
       \affaddr{University of Freiburg, Germany}\\
       \email{hornungt@informatik.uni-freiburg.de}
\and
\alignauthor Christopher Dorner\\
       \affaddr{University of Freiburg, Germany}\\
       \email{dornerc@informatik.uni-freiburg.de}
\alignauthor Georg Lausen\\
       \affaddr{University of Freiburg, Germany}\\
       \email{lausen@informatik.uni-freiburg.de}
}

\maketitle
\pagenumbering{arabic}

\begin{abstract}
One of the major challenges in large-scale data processing with MapReduce is the smart computation of joins. Since Semantic Web datasets published in RDF have increased rapidly over the last few years, scalable join techniques become an important issue for SPARQL query processing as well.
In this paper, we introduce the \emph{Map-Side Index Nested Loop Join} (MAPSIN join) which combines scalable indexing capabilities of NoSQL storage systems like HBase, that suffer from an insufficient distributed processing layer, with MapReduce, which in turn does not provide appropriate storage structures for efficient large-scale join processing.
While retaining the flexibility of commonly used reduce-side joins, we leverage the effectiveness of map-side joins without any changes to the underlying framework. We demonstrate the significant benefits of MAPSIN joins for the processing of SPARQL basic graph patterns on large RDF datasets by an evaluation with the LUBM and SP$^{2}$Bench benchmarks.
For most queries, MAPSIN join based query execution outperforms reduce-side join based execution by an order of magnitude.
\end{abstract}

\section{Introduction}
\label{sec-introduction}
Most of the information in the classical \emph{''Web of Documents''} is designed for human readers which makes it difficult for computers to automatically gather and process this data.
The idea behind the Semantic Web is to build a \emph{''Web of Data''} that enables computers to understand and use the information in the web \cite{bernerslee2001semantic}.
The advent of this Web of Data gives rise to new challenges with regard to query evaluation on the Semantic Web.
The core technologies of the Semantic Web are RDF (Resource Description Framework) for representing data in a machine-readable format and SPARQL for querying RDF data. In the last few years, more and more data has been published in RDF or other machine-readable formats. One of the most prominent examples is the \emph{Linking Open Data Cloud}\footnote{\url{http://www.w3.org/wiki/SweoIG/TaskForces/CommunityProjects/LinkingOpenData}}, a collection of interlinked RDF repositories with billions of RDF triples in total. Querying RDF datasets at web-scale is challenging, especially because the computation of SPARQL queries usually requires several joins between subsets of the data.
On the other side, classical single-place machine approaches have reached a point where they cannot scale with respect to the ever increasing amount of available RDF data (cf.~\cite{DBLP:journals/pvldb/HuangAR11}).
An intuitive solution is to parallelize the execution of query processing over a cluster of machines. Renowned for its excellent scaling properties, the MapReduce paradigm~\cite{dean2008mapreduce} is a good fit for such a distributed SPARQL engine. Following this avenue, we introduced the PigSPARQL project in~\cite{PigSPARQL}. PigSPARQL offers full support for SPARQL~1.0 and is implemented on top of the Apache Hadoop\footnote{\url{http://hadoop.apache.org}} and Pig\footnote{\url{http://pig.apache.org}} platform.

Subsequent evaluations of our prototype have shown that one of the major performance bottlenecks for SPARQL engines on MapReduce is the {\em computation of joins}. We have identified two major reasons for this shortcoming:

\begin{enumerate} 
  \item The Hadoop File System (HDFS) - a clone of the Google File System~\cite{ghemawat2003google} - provides no sufficient built-in index structures.
  \item A MapReduce-style join is typically computed during the reduce phase. As a side-effect, join partitions have to be sent across the network.
\end{enumerate}

In this paper we present the {\em Map-Side Index Nested Loop Join} (MAPSIN join), a completely map-side based join technique that uses HBase as underlying storage layer. Our evaluation shows an improvement of up to one order of magnitude over the common reduce-side join.

Overall, the major contributions of this paper are as follows:
\begin{itemize}
  \item We describe a space-efficient storage schema for large RDF graphs in HBase while retaining favourable access characteristics. By using HBase instead of HDFS, we can avoid shuffling join partitions across the network and instead only access the relevant join partners in each iteration.
  \item We present the MAPSIN join algorithm, which can be evaluated cascadingly in subsequent MapReduce iterations. In contrast to other approaches, we do not require an additional shuffle and reduce phase in order to preprocess the data for consecutive joins. Our approach solely relies on the complementary intertwining of a NoSQL data store and the MapReduce framework.
  \item We demonstrate an optimization of the basic MAPSIN join algorithm for the efficient processing of multiway joins. This way, we can save $n$ MapReduce iterations for star join queries with $n + 2$ triple patterns.
\end{itemize}

\noindent
The paper is structured as follows: Section~\ref{sec-foundations} provides a brief introduction to the technical foundations for this paper. Section~\ref{sec-rdfstorage} describes our RDF storage schema for HBase, while Section~\ref{sec-mapsidejoin} presents the MAPSIN join algorithm. We continue with a presentation of the evaluation of our approach in Section~\ref{sec-evaluation}, followed by a discussion of related work in Section~\ref{sec-relatedwork}. We conclude in Section~\ref{sec-conclusion} and give an outlook on future work.

\section{Background}
\label{sec-foundations}
In this chapter we briefly introduce the necessary core technologies used in our work, i.e. RDF, SPARQL, MapReduce, and HBase.
\subsection{RDF \& SPARQL}
\label{subsec:RDF_SPARQL}
\begin{figure}[htbp]
	\centering
	\includegraphics[scale=0.7]{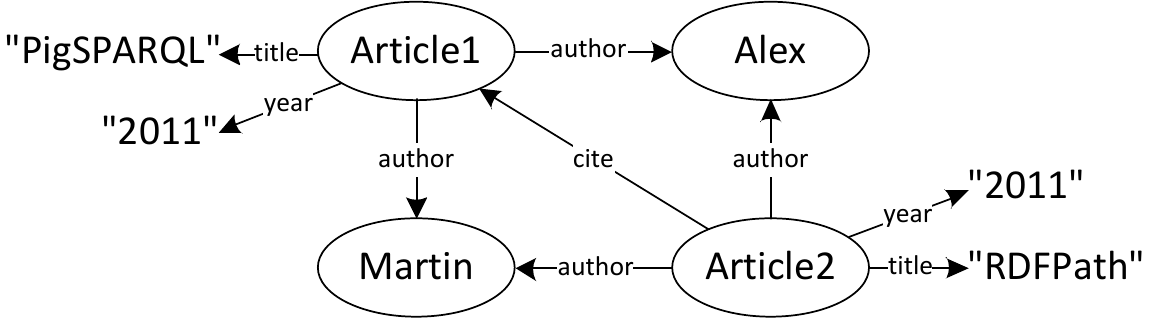}
	\caption{RDF graph}
	\label{fig:rdfgraph}
\end{figure}

RDF~\cite{rdfprimer} is the W3C recommended standard model for representing knowledge about arbitrary resources, e.g. articles and authors. An RDF dataset consists of a set of so-called RDF triples in the form (\emph{subject, predicate, object}) that can be interpreted as "\emph{subject} has property \emph{predicate} with value \emph{object}". For clarity of presentation, we use a simplified RDF notation in the following. It is possible to visualize an RDF dataset as directed, labeled graph where every triple corresponds to an edge (predicate) from subject to object. Figure \ref{fig:rdfgraph} shows an RDF graph with information about articles and corresponding authors.

SPARQL is the W3C recommended declarative query language for RDF. A SPARQL query defines a graph pattern $P$ that is matched against an RDF graph $G$. This is done by replacing the variables in $P$ with elements of $G$ such that the resulting graph is contained in $G$ (pattern matching). The most basic constructs in a SPARQL query are \emph{triple patterns}, i.e. RDF triples where subject, predicate and object can be variables, e.g. ($?s$, p, $?o$). A set of triple patterns concatenated by AND (.) is called a \emph{basic graph pattern} (BGP) as illustrated in the following query. The query asks for all articles with known title, author and year of publication. The result of a BGP is computed by joining the variable mappings of all triple patterns on their shared variables, in this case $?article$.

\begin{table}[htbp]
	\centering
	\label{tab:SPARQLexample}
		\begin{tabular}{l}
			\hline
			\textbf{Query 1:} \text{SPARQL Basic Graph Pattern Example}\\
			\hline
			\verb|SELECT *|\\
			\verb|WHERE {|\\
  			\verb|  ?article  title  ?title  .|\\
  			\verb|  ?article  author ?author .                     |\\
  			\verb|  ?article  year   ?year|\\
			\verb|}|
		\end{tabular}
\end{table}

There are other interesting SPARQL operators like FILTER and OPTIONAL that allow more sophisticated queries. For a detailed definition of the SPARQL syntax we refer the interested reader to the official W3C Recommendation~\cite{sparql}. A formal definition of the SPARQL semantics can also be found in~\cite{perez2009semantics}. In this paper we focus on efficient join processing with MapReduce and NoSQL (i.e. HBase) and therefore only consider SPARQL BGPs.
\subsection{MapReduce}
\label{subsec:MapReduce}

The MapReduce programming model was originally introduced by Google in 2004 \cite{dean2008mapreduce} and enables scalable, fault tolerant and massively parallel computations using a cluster of machines. The basis of Google's MapReduce is the distributed file system GFS \cite{ghemawat2003google} where large files are split into equal sized blocks, spread across the cluster and fault tolerance is achieved by replication.

\begin{figure}[htbp]
	\centering
		\includegraphics[scale=0.77]{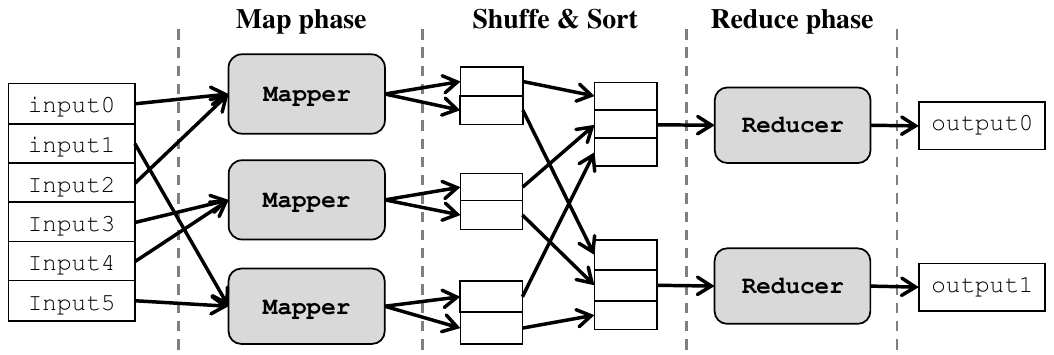}
	\caption{MapReduce Dataflow}
	\label{fig:mapreduce}
\end{figure}

The workflow of a MapReduce program is a sequence of MapReduce iterations each consisting of a \emph{Map} and a \emph{Reduce} phase separated by a so-called \emph{Shuffle} \& \emph{Sort} phase (cf. Figure \ref{fig:mapreduce}). The input data is automatically split in blocks and distributed over the cluster to enable parallel execution. A user has to implement \emph{map} and \emph{reduce} functions which are automatically executed in parallel on a portion of the data. The Mappers invoke the map function for every record of their input dataset represented as a key-value pair. The map function outputs a list of new intermediate key-value pairs which are then sorted according to their key and distributed to the Reducers such that all values with the same key are sent to the same Reducer. The reduce function is invoked for every distinct key together with a list of all according values and outputs a list of values which can be used as input for the next MapReduce job. The signatures of the map and reduce functions are therefore as follows:

\begin{verbatim}
map:    (inKey, inValue) -> list(outKey, tmpValue)
reduce: (outKey, list(tmpValue)) -> list(outValue)
\end{verbatim}

\medskip
We use \emph{Hadoop}~\cite{hadoopGuideWhite} as it is the most popular open-source implementation of Google's GFS and MapReduce framework that is used by many companies like Yahoo!, IBM or Facebook. Hadoop is enriched with an extensive ecosystem of related projects that are either built on top of Hadoop or use Hadoop for various application fields.

\paragraph{Map-Side vs. Reduce-Side Join}
Processing joins with MapReduce is a challenging task as datasets are typically very large~\cite{DBLP:conf/sigmod/BlanasPERST10,DBLP:journals/sigmod/LeeLCCM11}.
If we want to join two datasets with MapReduce, $L \Join R$, we have to ensure that the subsets of $L$ and $R$ with the same join key values can be processed on the same machine.
For joining arbitrary datasets on arbitrary keys we generally have to shuffle data over the network or choose appropriate pre-partitioning and replication strategies.

The most prominent and flexible join technique in MapReduce is called \emph{Reduce-Side Join}~\cite{DBLP:conf/sigmod/BlanasPERST10,DBLP:journals/sigmod/LeeLCCM11,2010Lin,hadoopGuideWhite}.
Some literature also refer to it as \emph{Repartition Join}~\cite{DBLP:conf/sigmod/BlanasPERST10} as the basic idea is based on reading both datasets (map phase) and repartition them according to the join key (shuffle phase). The actual join computation is done in the reduce phase. The main drawback of this approach is that both datasets are completely transferred over the network regardless of the join output. This is especially inefficient for selective joins and consumes a lot of network bandwidth.
 
Another group of joins is based on getting rid of the shuffle and reduce phase to avoid transferring both datasets over the network. This kind of join technique is called \emph{Map-Side Join} since the actual join processing is done in the map phase~\cite{hadoopGuideWhite}. The most common one is the \emph{Map-Side Merge Join}~\cite{DBLP:journals/sigmod/LeeLCCM11,2010Lin}. However, such a join cannot be applied on arbitrary datasets. A preprocessing step is necessary to fulfill several requirements: datasets have to be sorted and equally partitioned according to the join key. If the preconditions are fulfilled, the map phase can process an efficient parallel merge join between pre-sorted partitions and data shuffling is not necessary. However, if we want to compute a sequence of joins, the shuffle and reduce phases are needed to guarantee that the preconditions for the next join iteration are fulfilled.
Therefore, map-side joins are generally hard to cascade and the advantage of avoiding a shuffle and reduce phase is lost. Our MAPSIN join approach is designed to overcome this drawback by using the distributed index of a NoSQL storage system like HBase.
\subsection{HBase}
HBase~\cite{george2011hbase} is a distributed, scalable and strictly consistent column-oriented NoSQL storage system, inspired by Google's Bigtable \cite{chang2008bigtable} and well integrated into Hadoop, the most popular open-source implementation of Google's MapReduce framework. Hadoop's distributed file system, HDFS, is designed for sequential reads and writes of very large files in a batch processing manner but lacks the ability to access data randomly in close to real-time \cite{george2011hbase}. HBase can be seen as an additional storage layer on top of HDFS that supports efficient random access.
The data model of HBase corresponds to a sparse multi-dimensional sorted map with the following access pattern:
\[(Table, Row Key, Family, Column, Timestamp) \rightarrow Value\]
Data is stored in tables that are structurally different from tables in relational databases. The rows of a table are sorted according to their \emph{row key} and every row can have an arbitrary number of \emph{columns}. However, in contrast to relational tables, two distinct rows of an HBase table do not have to have the same columns. Columns are grouped into \emph{column families} that must be defined at the time of table creation and should not change too often whereas columns can be added and deleted dynamically as needed. Column values (denoted as cell) are timestamped and thus support multiple versions where the user can specify how many versions should be stored.

HBase tables are dynamically split into \emph{regions} of contiguous row ranges with a configured maximum size. When a region becomes too large, it is automatically split into two regions at the middle key (auto-sharding). Each region is assigned to exactly one so-called \emph{region server} in an HBase cluster whereas a region server can serve many regions. All columns of a column family within a region are stored together in the same low-level storage file in HDFS. As these files are ordered and indexed by row key, table lookups and range scans based on row key can be done efficiently.

However, HBase has neither a declarative query language like SQL nor a built-in support for native join processing, leaving higher-level data transformations to the overlying application layer.
In our approach we propose a map-side join strategy that leverages the implicit index capabilities of HBase to overcome the usual restrictions of map-side joins as outlined in Section \ref{subsec:MapReduce}.

\section{RDF Storage Schema for HBase}
\label{sec-rdfstorage}
In contrast to relational databases, NoSQL storage systems do not have a common data model in general. This makes it impossible to define a universal storage schema for RDF that holds for any kind of NoSQL system. Furthermore, there is no common query language like SQL and many systems do not have any query language at all. Hence, the implementation of our join approach strongly relies on the actual NoSQL system used as backend. In our initial experiments we considered HBase and Cassandra~\cite{hewitt2010cassandra}, two popular NoSQL systems with support for MapReduce. We decided to use HBase for our implementation as it proved to be more stable and also easier to handle in our cluster since HBase was developed to work with Hadoop from the beginning.

In \cite{sun2010} the authors adopted the idea of Hexastore \cite{weiss2008} to index all possible orderings of an RDF triple for storing RDF data in HBase. This results in six tables in HBase allowing to retrieve results for any possible SPARQL triple pattern with a single GET on one of the tables (except for a triple pattern with three variables). However, as HDFS has a default replication factor of three and data in HBase is stored in files on HDFS, an RDF dataset is actually stored 18 times using this schema. But it's not only about storage space, also loading a web-scale RDF dataset into HBase becomes very costly and consumes many resources.

Our storage schema for RDF data in HBase is inspired by \cite{franke2011} and uses only two tables, $T_{s\_po}$ and $T_{o\_ps}$. We extend the schema with a triple pattern mapping that leverages the power of predicate push-down filters in HBase to compensate possible performance shortcomings of a two table schema compared to the six table schema. Furthermore, we improve the scalibility of the schema by introducing a modified row key design for class assignments in RDF which would otherwise lead to overloaded regions constraining both scalability and performance.

In $T_{s\_po}$ table an RDF triple is stored using the subject as row key, the predicate as column name and the object as column value. If a subject has more than one object for a given predicate (e.g. an article having more than one author), these objects are stored as different versions in the same column. The notation $T_{s\_po}$ indicates that the table is indexed by subject. Table $T_{o\_ps}$ follows the same design. In both tables there is only one single column family that contains all columns.
Tables~\ref{tab:TSP} and \ref{tab:TOP} illustrate how the RDF graph in Section \ref{subsec:RDF_SPARQL} would be represented in the HBase storage schema.

\begin{table}[htbp]
	\centering
	\caption{$T_{s\_po}$ table for RDF graph in Section \ref{subsec:RDF_SPARQL}}
	\label{tab:TSP}
		\begin{tabularx}{\columnwidth}{|X|l|}
			\hline
			\textbf{rowkey}    &   \textbf{family:column}$\rightarrow$\textbf{value}\\
			\hline
			Article1           &   p:title$\rightarrow$\{''PigSPARQL''\}, p:year$\rightarrow$\{''2011''\},\\
			                   &   p:author$\rightarrow$\{Alex, Martin\}\\
			\hline
			Article2           &   p:title$\rightarrow$\{''RDFPath''\}, p:year$\rightarrow$\{''2011''\},\\
			                   &   p:author$\rightarrow$\{Martin, Alex\}, p:cite$\rightarrow$\{Article1\}\\
			\hline
		\end{tabularx}
\end{table}

\begin{table}[htbp]
	\centering
	\caption{$T_{o\_ps}$ table for RDF graph in Section \ref{subsec:RDF_SPARQL}}
	\label{tab:TOP}
		\begin{tabularx}{\columnwidth}{|l|X|}
			\hline
			\textbf{rowkey}    &   \textbf{family:column}$\rightarrow$\textbf{value}\\
			\hline
			''2011''           &   p:year$\rightarrow$\{Article1, Article2\}\\
			\hline
			''PigSPARQL''      &   p:title$\rightarrow$\{Article1\}\\
			\hline
			''RDFPath''        &   p:title$\rightarrow$\{Article2\}\\
			\hline
			Alex               &   p:author$\rightarrow$\{Article1, Article2\}\\
			\hline
			Article1           &   p:cite$\rightarrow$\{Article2\}\\
			\hline
			Martin             &   p:author$\rightarrow$\{Article2, Article1\}\\
			\hline
		\end{tabularx}
\end{table}

At first glance, this storage schema seems to have some serious performance drawbacks when compared to the six table schema in \cite{sun2010} since there are only indexes for subjects and objects. In the six table schema there is, for example, also an index for the combination of subject and predicate, resulting in a simple GET on the corresponding table for a triple pattern with bounded subject and predicate. Using $T_{s\_po}$ table, we can only make a GET for the given subject and filter the result for the given predicate. However, we can use the HBase Filter API to specify an additional column filter for the GET request. This filter is applied directly on server side such that no unnecessary data must be transferred over the network (\emph{predicate push-down}).

As already mentioned in \cite{franke2011}, having a table with predicates as row keys has some serious scalability problems since the number of predicates in an ontology is usually fixed and relatively small which results in a table with just a few very fat rows. Considering that all data in a row is stored on the same machine, the resources of a single machine in the cluster become a bottleneck. Indeed, if only the predicate in a triple pattern is given, we can use the HBase Filter API to answer this request with a table SCAN on $T_{s\_po}$ or $T_{o\_ps}$ using the predicate as column filter.
Table \ref{tab:Mapping} shows the mapping of every possible triple pattern to the corresponding HBase table. Overall, experiments on our cluster showed that the two table schema with server side filters has similar performance characteristics compared to the six table schema but uses only one third of storage space.

\begin{table}[htbp]
	\centering
	\caption{SPARQL triple pattern mapping using HBase predicate push-down filters}
	\label{tab:Mapping}
		\begin{tabularx}{\columnwidth}{|X|l|l|}
			\hline
			\textbf{pattern}      &   \textbf{table}                    	    &   \textbf{filter}\\
			\hline
			(s, p, o)             &   $T_{s\_po}$ or $T_{o\_ps}$                &   column \& value\\
			\hline
			(?s, p, o)            &   $T_{o\_ps}$                          	    &   column\\
			\hline
			(s, ?p, o)            &   $T_{s\_po}$ or $T_{o\_ps}$                &   value\\
			\hline
			(s, p, ?o)            &   $T_{s\_po}$                               &   column\\
			\hline
			(?s, ?p, o)           &   $T_{o\_ps}$                               &   \\
			\hline
			(?s, p, ?o)           &   $T_{s\_po}$ or $T_{o\_ps}$ (SCAN)         &   column\\
			\hline
			(s, ?p, ?o)           &   $T_{s\_po}$                               &   \\
			\hline
			(?s, ?p, ?o)          &   $T_{s\_po}$ or $T_{o\_ps}$ (SCAN)         &   \\
			\hline
		\end{tabularx}
\end{table}

Our experiments revealed some serious scaling limitations of the storage schema caused by the $T_{o\_ps}$ table. In general, an RDF dataset uses a relatively small number of classes but contains many triples that link resources to classes, e.g. (Alex, rdf:type, foaf:Person). Thus, using the object of a triple as row key means that all resources of the same class will be stored in the same row. With increasing dataset size these rows become very large and exceed the configured maximum region size resulting in overloaded regions that contain only a single row. Since HBase cannot split these regions the resources of a single machine become a bottleneck for scalability. To circumvent this problem we use a modified $T_{o\_ps}$ row key design for triples with predicate rdf:type. Instead of using the object as row key we use a compound row key of object and subject, e.g. (foaf:Person$|$Alex). As a result, we can not access all resources of a class with a single GET but as the corresponding rows of a class will be consecutive in $T_{o\_ps}$ we can use an efficient range SCAN starting at the first entry of the class.

\section{MAPSIN Join}
\label{sec-mapsidejoin}
The major task in SPARQL query evaluation is the computation of joins between triple patterns, i.e.~a basic graph pattern. However, join processing on large RDF datasets, especially if it involves more than two triple patterns, is challenging~\cite{DBLP:journals/sigmod/LeeLCCM11}.
Section \ref{subsec:MapReduce} discussed the most common join techniques in MapReduce and stated the importance of reducing the amount of data transferred over the network as well as typical drawbacks of map-side join approaches.

Our approach combines the scalable storage capabilities of NoSQL datastores (i.e. HBase), that suffer from a suitable distributed processing layer, with MapReduce, a highly scalable and distributed computation framework, which in turn does not support appropriate storage structures for large scale join processing. This allows us to catch up with the flexibility of reduce-side joins while utilizing the effectiveness of a map-side join that does not need a shuffle and reduce phase without any changes to the underlying Hadoop framework.\\

\noindent
First, we introduce the needed SPARQL terminology analogous to~\cite{perez2009semantics}: Let $V$ be the infinite set of query variables and $T$ the set of valid RDF terms.

\newdef{definition}{Definition}
\begin{definition}
A (solution) mapping $\mu$ is a partial function $\mu:V \rightarrow T$. We call $\mu(?v)$ the variable binding of $\mu$ for $?v$. Abusing notation, for a triple pattern $p$ we call $\mu(p)$ the triple pattern that is obtained by substituting the variables in $p$ according to $\mu$. The domain of $\mu$, $dom(\mu)$, is the subset of $V$ where $\mu$ is defined and the domain of $p$, $dom(p)$, is the subset of $V$ used in $p$. The result of a SPARQL query is a multiset of solution mappings $\Omega$.
\end{definition}

\begin{definition}
Two solution mappings $\mu_1, \mu_2$ are compatible if, for every variable $?v \in dom(\mu_1) \cap dom(\mu_2)$, it holds that $\mu_1(?v) = \mu_2(?v)$. It follows that solution mappings with disjoint domains are always compatible and the set-union (merge) of $\mu_1$ and $\mu_2$, $\mu_1 \cup \mu_2$, is also a solution mapping.
\end{definition}

\subsection{Base Case}
\label{subsec:BaseCase}
Let us assume our RDF graph is stored in a NoSQL storage system like HBase as described in Section~\ref{sec-rdfstorage} and therefore spread across a cluster of machines. If we want to process a basic graph pattern of two concatenated triple patterns $p_1$ and $p_2$, we have to compute the join $p_1 \Join p_2$ that merges the compatible mappings of two multisets of solution mappings. Hence, it is necessary that subsets of both multisets of mappings are brought together such that it is ensured that all compatible mappings can be processed on the same machine.

Our MAPSIN join technique computes the join between $p_1$ and $p_2$ in a single map phase. At the beginning, the map phase is initialized with a parallel distributed HBase table scan for the first triple pattern $p_1$ where each machine retrieves only those mappings that are locally available. This is achieved by utilizing a mechanism for allocating local records to map functions, which is supported by the MapReduce input format for HBase automatically. 

The map function is invoked for each retrieved mapping $\mu_{1}$ for $p_1$. To compute the partial join between $p_1$ and $p_2$ for the given mapping $\mu_{1}$, the map function needs to retrieve those mappings for $p_2$ that are compatible to $\mu_{1}$ based on the shared variables between $p_1$ and $p_2$. At this point, the map function utilizes the input mapping $\mu_{1}$ to substitute the shared variables in $p_2$, i.e. the join variables. Now, the substituted triple pattern $p_2^{sub}$ is used to retrieve the compatible mappings with a dynamic GET request to HBase following the triple pattern mapping outlined in Table~\ref{tab:Mapping}. Since there is no guarantee that the corresponding HBase entries reside on the same machine as the GET request, the results of the request have to be transferred over the network in general. However, in contrast to a reduce-side join approach where a lot of data is transferred over the network, we only transfer the data that is really needed. Finally, the computed multiset of solution mappings is stored in HDFS.

\begin{figure}[htpb]
	\centering
	\includegraphics[scale=0.56]{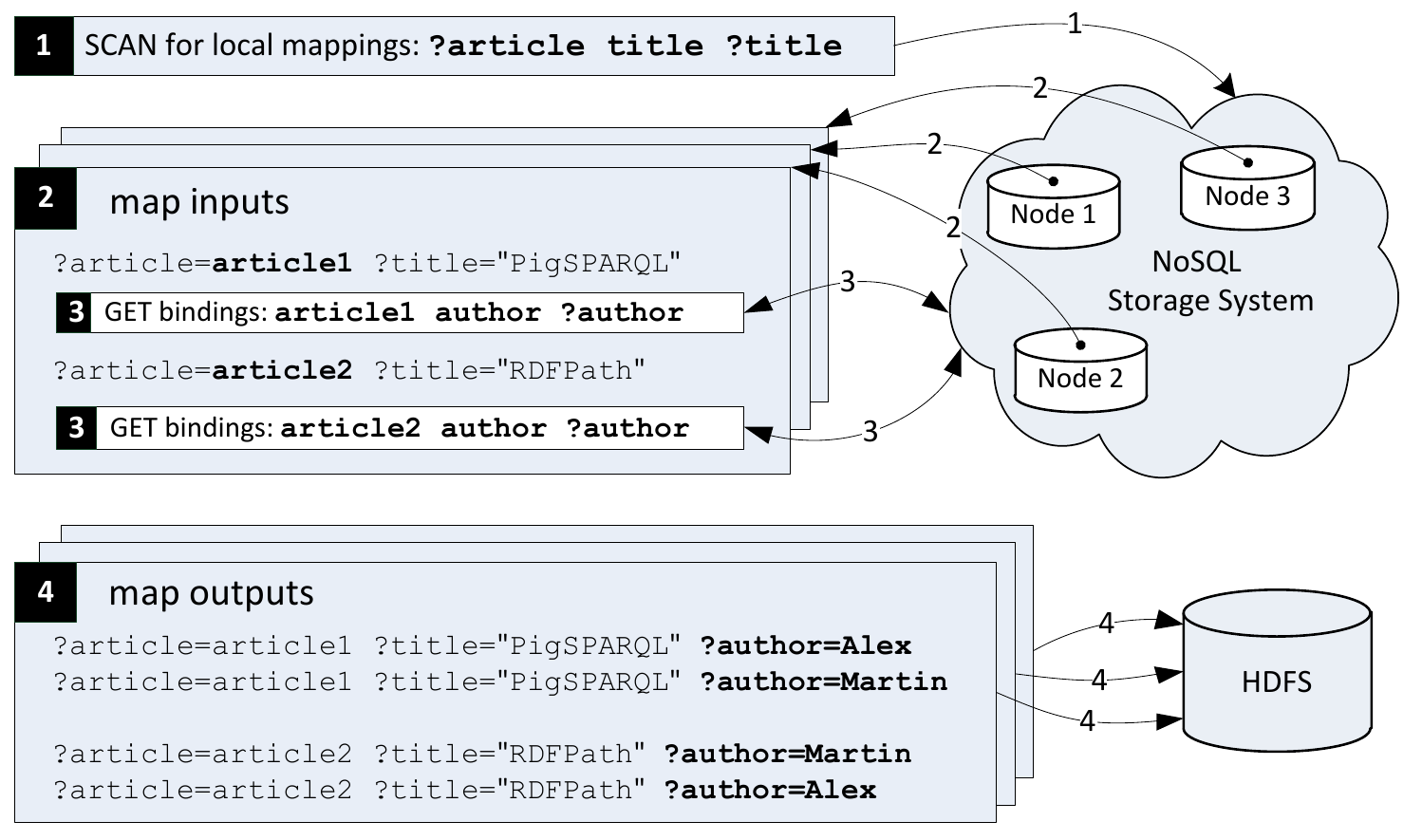}
	\caption{MAPSIN join base case for the first two triple patterns of Query 1 in Section~\ref{subsec:RDF_SPARQL}}
	\label{fig:map-side-join-example}
\end{figure}

Figure \ref{fig:map-side-join-example} is an example for the base case of our MAPSIN join that illustrates the join between the first two triple patterns of the SPARQL query in Section~\ref{subsec:RDF_SPARQL}. While the mappings for the first triple pattern ($?article$, title, $?title$) are retrieved locally using a distributed table scan (step 1+2), the compatible mappings for ($?article$, author, $?author$) are requested within the map function dynamically (step 3) and the resulting set of mappings is stored in HDFS (step 4).

\subsection{Cascading Joins}
\label{subsec:CascadingJoins}
Chains of concatenated triple patterns require some slight modifications to the previously described base case. To compute a query of at least three triple patterns we have to process several joins successively, e.g. $p_1 \Join p_2 \Join p_3$. The processing of the first two patterns $p_1 \Join p_2$ correspond to the base case and the results are stored in HDFS. The additional triple pattern $p_3$ is then joined with the solution mappings of $p_1 \Join p_2$. To this end, an additional map-phase (without any intermediate shuffle or reduce phase) is initialized with the previously computed solution mappings as input. Since these mappings reside in HDFS, they are retrieved locally in parallel such that the map function gets invoked for each solution mapping $\mu_{2}$ of $p_1 \Join p_2$. The compatible mappings for $p_3$ are retrieved using the same strategy as for the base case, i.e. $\mu_{2}$ is used to substitute the shared variables in $p_3$ and compatible mappings are retrieved following the triple pattern mapping outlined in Table~\ref{tab:Mapping}.

We implemented a basic heuristic called \emph{variable counting} to reorder triple patterns in a SPARQL BGP by selectivity based on the number and position of variables in a pattern~\cite{stocker2008}. The heuristic assumes that triple patterns with one variable are more selective than triple patterns with two variables and bounded subjects are more selective than bounded predicates or objects. This reduces the number of intermediate results in a sequence of joins and thus reduce the number of requests against HBase.

\begin{algorithm}[htbp]
\DontPrintSemicolon
\SetKwInOut{Input}{input}
\SetKwInOut{Output}{output}
\Input{		\emph{inKey},
			\emph{inValue:} input mapping}
\Output{	multiset of mappings}
		
$p_{n+1} \leftarrow$ Config.getNextPattern()\;
$\mu_{n} \leftarrow$ $inValue$.getInputMapping()\;
$\Omega_{n+1} \leftarrow \emptyset$\;

\eIf{$dom(\mu_{n}) \cap dom(p_{n+1}) \neq \emptyset$}{	
	// substitute shared vars in $p_{n+1}$\;
	$p_{n+1}^{sub} \leftarrow \mu_{n}(p_{n+1})$\;
	$results \leftarrow$ HBase.GET($p_{n+1}^{sub}$)\;
}{
	$results \leftarrow$ HBase.GET($p_{n+1}$)\;
}
\If{$results \neq \emptyset$}{
	// merge $\mu_{n}$ with compatible mappings for $p_{n+1}$\;
	\ForEach{mapping $\mu$ in $results$}{
		$\mu_{n+1} \leftarrow \mu_{n} \cup \mu$\;
		$\Omega_{n+1} \leftarrow \Omega_{n+1} \cup \mu_{n+1}$\;
	}
	emit($null$, $\Omega_{n+1}$)\;
}
\caption{MAPSIN join \textbf{map}(inKey, inValue)}
\label{algorithm1}
\end{algorithm}

Algorithm \ref{algorithm1} outlines one iteration of the MAPSIN join (without triple pattern reordering). The input for the map function contains either a mapping for the first triple pattern or a mapping for previously joined triple patterns. The algorithm illustrates why we refer to our join technique as a map-side index nested loop join. First of all, the join is completely computed in the map phase. Furthermore, the map function is invoked for each input mapping what corresponds to an outer loop. Within the map function, we retrieve the compatible mappings for the next triple pattern using index lookups in HBase.

\subsection{Multiway Join Optimization}
\label{subsec:MAPSIN_MULTI}
Instead of processing concatenated triple patterns successively as a sequence of two-way joins, some basic graph patterns allow to apply a multiway join approach to process joins between several concatenated triple patterns at once in a single map phase. This is typically the case for star pattern queries where triple patterns share the same join variable. The SPARQL query introduced in Section~\ref{subsec:RDF_SPARQL} is an example for such a query as all triple patterns share the same join variable $?article$. This query can be processed by a three-way join in a single map-phase instead of two consecutive two-way joins.

We extended our approach to support this multiway join optimization. Again, the first triple pattern $p_{1}$ is processed using a distributed table scan as input for the map phase. But instead of using a sequence of $n$ map phases to compute $p_{1}\Join p_{2}\Join ...\Join p_{n+1}$ we use a single map phase thus saving $n-1$ MapReduce iterations. Hence, the map function needs to retrieve all mappings for $p_{2},p_{3},...,p_{n+1}$ that are compatible to the input mapping $\mu_{1}$ for $p_1$. Therefore, the join variable $?v_{s}$ in $p_{2},p_{3},...,p_{n+1}$ (e.g. $?article$) is substituted with the corresponding variable binding $\mu(?v_{s})$. The substituted triple patterns $p_{2}^{sub},p_{3}^{sub},...,p_{n+1}^{sub}$ are then used to retrieve the compatible mappings using HBase GET requests. This general case of the MAPSIN multiway join is outlined in Algorithm \ref{algorithm2}.

\begin{algorithm}[htbp]
\DontPrintSemicolon
\SetKwInOut{Input}{input}
\SetKwInOut{Output}{output}
\Input{		\emph{inKey},
			\emph{inValue:} input mapping}
\Output{	multiset of mappings}
		
$\#p \leftarrow $ Config.getNumberOfMultiwayPatterns()\;
$\mu_{n} \leftarrow$ $inValue$.getInputMapping()\;
$\Omega_{n} \leftarrow \{\mu_{n}\}$\;

// iterate over all subsequent multiway patterns\;
\For{$i\leftarrow 1$ \KwTo $\#p$}{
	$\Omega_{n+i} \leftarrow \emptyset$\;
	$p_{n+i} \leftarrow$ Config.getNextPattern()\;
	// substitute shared vars in $p_{n+i}$\;
	$p_{n+i}^{sub} \leftarrow \mu_{n}(p_{n+i})$\;
	$results \leftarrow$ HBase.GET($p_{n+i}^{sub}$)\;
	\eIf{$results \neq \emptyset$}{
		// merge previous mappings with\;
		// compatible mappings for $p_{n+i}$\;
		\ForEach{mapping $\mu$ in $results$}{
			\ForEach{mapping $\mu'$ in $\Omega_{n+i-1}$}{
				$\Omega_{n+i} \leftarrow \Omega_{n+i} \cup (\mu \cup \mu')$\;
			}
		}	
	}{
		// no compatible mappings for $p_{n+i}$\;
		// hence join result for $\mu_{n}$ is empty\;
		return\;
	}
}
emit($null$, $\Omega_{n+\#p}$)\;
\caption{MAPSIN multiway join \newline \textbf{map}(inKey, inValue)}
\label{algorithm2}
\end{algorithm}

In many situations, it is possible to reduce the amount of requests against HBase by leveraging the RDF schema design for HBase outlined in Section~\ref{sec-rdfstorage}. If the join variable for all triple patterns is always on subject or always on object position, then all mappings for $p_{2},p_{3},...,p_{n+1}$ that are compatible to the input mapping $\mu_{1}$ for $p_1$ are stored in the same HBase table row of $T_{s\_po}$ or $T_{o\_ps}$, respectively, making it possible to use a single instead of $n$ subsequent GET requests. Hence, all compatible mappings can be retrieved at once thus saving $n-1$ GET requests for each invocation of the map function. Algorithm \ref{algorithm3} outlines this optimized case.

\begin{algorithm}[htbp]
\DontPrintSemicolon
\SetKwInOut{Input}{input}
\SetKwInOut{Output}{output}
\Input{		\emph{inKey},
			\emph{inValue:} input mappings}
\Output{	multiset of solution mappings}
		
$\#p \leftarrow $ Config.getNumberOfMultiwayPatterns()\;
$\mu_{n} \leftarrow$ $inValue$.getInputMapping()\;
$\Omega_{n+\#p} \leftarrow \emptyset$\;

// iterate over all subsequent multiway patterns\;
\For{$i\leftarrow 1$ \KwTo $\#p$}{
	$p_{n+i} \leftarrow$ Config.getNextPattern()\;
	// substitute shared vars in $p_{n+i}$\;
	$p_{n+i}^{sub} \leftarrow \mu_{n}(p_{n+i})$\;
}
$results \leftarrow$ HBase.GET($p_{n+1}^{sub}$, ..., $p_{n+\#p}^{sub}$)\;
\If{$results \neq \emptyset$}{
	// merge $\mu_{n}$ with compatible\;
	// mappings for $p_{n+1}, ..., p_{n+\#p}$\;
	\ForEach{mapping $\mu$ in $results$}{
		$\mu_{n+\#p} \leftarrow \mu_{n} \cup \mu$\;
		$\Omega_{n+\#p} \leftarrow \Omega_{n+\#p} \cup \mu_{n+\#p}$\;
	}
	emit($null$, $\Omega_{n+\#p}$)\;
}
\caption{MAPSIN optimized multiway join \newline \textbf{map}(inKey, inValue)}
\label{algorithm3}
\end{algorithm}

\subsection{One Pattern Queries}
Queries with only one single triple pattern, that return only a small number of solution mappings, can be executed locally on one machine. In general however, the number of solution mappings can exceed the capabilities of a single machine for large datasets. Thus, concerning scalability it is advantageous to use a distributed execution with MapReduce even if we do not need to perform a join.

The map phase is initialized with a distributed table scan for the single triple pattern $p_1$. Hence, map functions get only those mappings as input, which are locally available. The map function itself has only to emit the mappings to HDFS without any further requests to HBase. If the query result is small, non-distributed query execution can reduce query execution time significantly as MapReduce initialization takes up to 30 seconds in our cluster which clearly dominates the execution time.

\definecolor{gray}{gray}{0.95}
\section{Evaluation}
\label{sec-evaluation}
The evaluation was performed on a cluster of 10 Dell PowerEdge R200 servers equipped with a Dual Core 3.16 GHz CPU, 8 GB RAM, 3 TB disk space and connected via gigabit network.
The software installation includes Hadoop 0.20.2, HBase 0.90.4 and Java 1.6.0 update 26.

\begin{table*}[htbp]
	\centering
	\caption{SP$^2$Bench \& LUBM loading times for tables $T_{s\_po}$ and $T_{o\_ps}$ (hh:mm:ss)}
	\label{tab:Loading}
		\begin{tabularx}{\textwidth}{|X|c|c|c|c|c|}
			\hline			
			\textbf{SP$^2$Bench}   &   \textbf{200M} &  \textbf{400M} &  \textbf{600M} &  \textbf{800M} &  \textbf{1000M}\\
			\hline
			\# RDF triples     	   &   $\sim$ 200 million      &  $\sim$ 400 million      &  $\sim$ 600 million      &  $\sim$ 800 million      &  $\sim$ 1000 million\\
			\hline
			$T_{s\_po}$            &   ~~~~~~00:28:39~~~~~~  &  ~~~~~~00:45:33~~~~~~  &  ~~~~~~01:01:19~~~~~~  &  ~~~~~~01:16:09~~~~~~  &  ~~~~~01:33:47~~~~~~\\
			\hline
			$T_{o\_ps}$            &   00:27:24      &  01:04:30      &  01:28:23      &  01:43:36      &  02:19:05\\
			\hline
			total           	   &   00:56:03      &  01:50:03      &  02:29:42      &  02:59:45      &  03:52:52\\
			\hline
			\hline			
			\textbf{LUBM}          &   \textbf{1000} &  \textbf{1500} &  \textbf{2000} &  \textbf{2500} &  \textbf{3000}\\
			\hline
			\# RDF triples     	   &   $\sim$ 210 million      &  $\sim$ 315 million      &  $\sim$ 420 million      &  $\sim$ 525 million      &  $\sim$ 630 million\\
			\hline
			$T_{s\_po}$            &   00:28:50      &  00:42:10      &  00:52:03      &  00:56:00      &  01:05:25\\
			\hline
			$T_{o\_ps}$            &   00:48:57      &  01:14:59      &  01:21:53      &  01:38:52      &  01:34:22\\
			\hline
			total           	   &   01:17:47      &  01:57:09      &  02:13:56      &  02:34:52      &  02:39:47\\
			\hline
		\end{tabularx}
\end{table*}

Most of the queries are taken from the well-known Lehigh University Benchmark (LUBM)~\cite{guo2005lubm} as this benchmark is a well-suited choice when considering join processing in a Semantic Web scenario since the queries of the benchmark can easily be formulated as SPARQL basic graph patterns.
Furthermore, we also considered the SPARQL-specific SP$^2$Bench Performance Benchmark \cite{schmidt2009sp}. However, because most of the SP$^2$Bench queries are rather complex queries that use all different kinds of SPARQL 1.0 operators, we only evaluated some of the queries as the focus of our work is the efficient computation of joins, i.e. SPARQL basic graph patterns. We decided to use both benchmarks as they have different characteristics and LUBM queries are generally more selective than SP$^2$Bench queries.
Both benchmarks offer synthetic data generators that can be used to generate arbitrary large datasets. For SP$^2$Bench we generated datasets from 200 million up to 1000 million triples. For LUBM we generated datasets from 1000 up to 3000 universities and used the WebPIE inference engine for Hadoop \cite{urbani2010owl} to pre-compute the transitive closure. For loading data into HBase, we set the region size to 512 MB using snappy compression. The loading times for both tables $T_{s\_po}$ and $T_{o\_ps}$ as well as all datasets are listed separately in Table \ref{tab:Loading}.

We compared our MAPSIN join approach with a reduce-side join based query execution using PigSPARQL \cite{PigSPARQL}, a SPARQL 1.0 engine built on top of \emph{Pig}. Pig is an Apache top-level project developed by Yahoo! Research that offers a high-level language for the analysis of very large datasets with Hadoop MapReduce. The crucial point for this choice was the sophisticated and efficient reduce-side join implementation of Pig~\cite{gates2009} that incorporates sampling and hash join techniques which makes it a challenging candidate for comparison. Pig also supports map-side joins as outlined in Section \ref{subsec:MapReduce}. However, these joins cannot be easily cascaded as the preconditions for the next iteration are not fulfilled without a join postprocessing using shuffle and reduce phase.

In order to avoid caching effects that could possibly influence our experiments, HBase was restarted before each query execution.
For detailed comparison, all query execution times are listed in Table \ref{tab:ExecutionTimes}. We illustrate the performance comparison of PigSPARQL and MAPSIN for some selected queries that represent the different query types. For completeness, all query evaluations can be found in Appendix \ref{appendix-lubm} for LUBM and Appendix \ref{appendix-sp2bench} for SP$^2$Bench.\\

\begin{figure}[h]
	\centering
		\begin{tabularx}{\columnwidth}{|X|}
			\hline
			\rowcolor{gray} \textbf{LUBM Q1.} \textit{Return all graduate students that attend a given course from a given university department}\\
			\hline
			\scriptsize{\verb|SELECT ?X|}\\
			\scriptsize{\verb|WHERE | \{}\\
			\scriptsize{~~\verb|?X rdf:type ub:GraduateStudent.|}\\
			\scriptsize{~~\verb|?X ub:takesCourse|}\\
			\scriptsize{~~~~~~\verb|<http://www.Department0.University0.edu/GraduateCourse0>| \}}\\
			\includegraphics[scale=0.67]{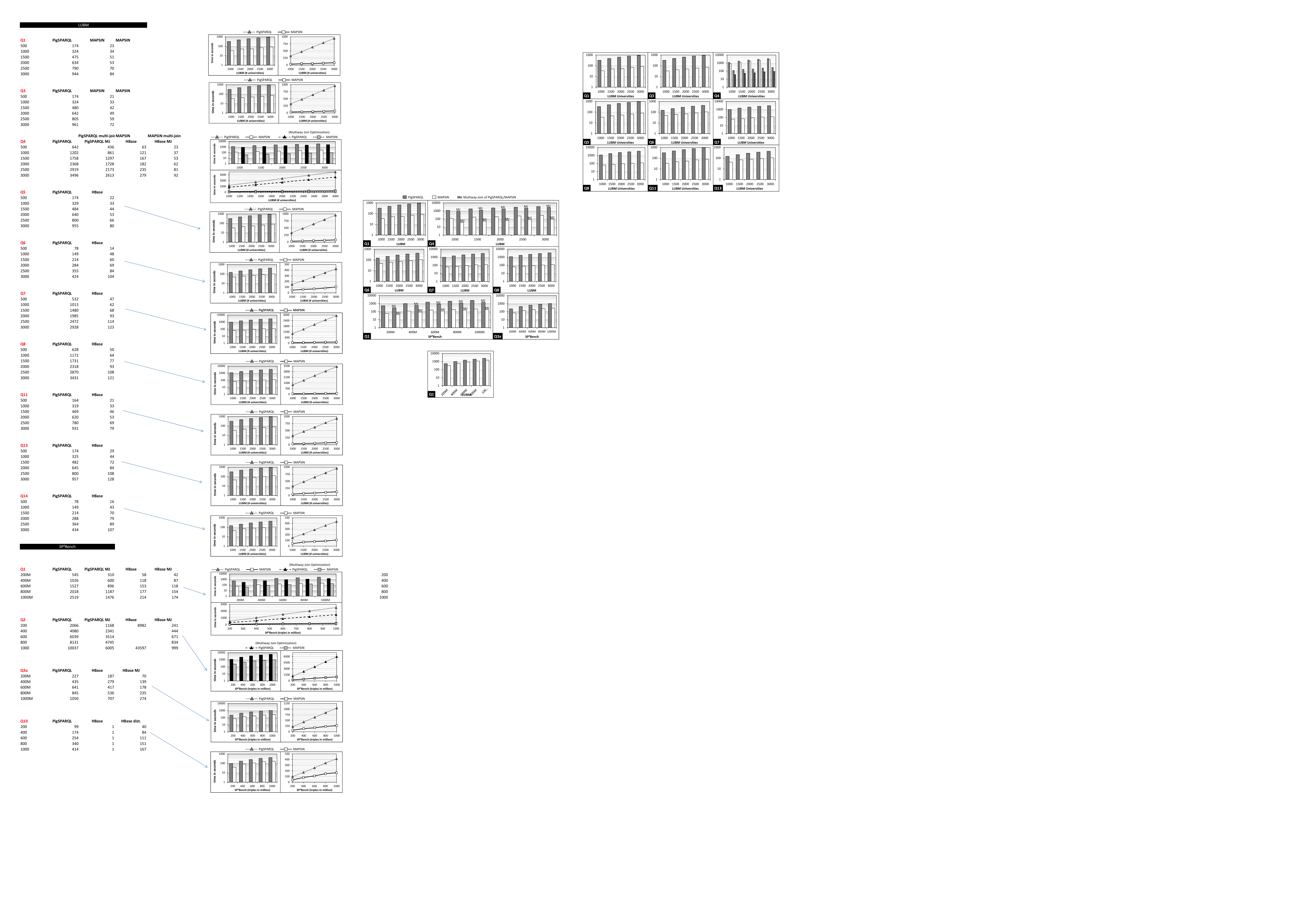}\\
			\hline	
		\end{tabularx}
	\caption{Performance comparison for LUBM Q1}
	\label{fig:eval_q1}
\end{figure}

LUBM queries Q1, Q3, Q5, Q11, Q13 as well as SP$^2$Bench query Q3a demonstrate the base case with a single join between two triple patterns (cf. Figure~\ref{fig:eval_q1}). For the LUBM queries, MAPSIN joins performed 8 to 13 times faster compared to the reduce-side joins used by PigSPARQL. Even for the less selective SP$^2$Bench query, our MAPSIN join required only one third of the PigSPARQL execution time. Furthermore, the performance gain increases with the size of the dataset for both LUBM and SP$^2$Bench.

\begin{figure}[h]
	\centering
		\begin{tabularx}{\columnwidth}{|X|}
			\hline
			\rowcolor{gray} \textbf{LUBM Q8.} \textit{Return the email addresses of all students that are members of any department of a given university}\\
			\hline
			\scriptsize{\verb|SELECT ?X, ?Y, ?Z|}\\
			\scriptsize{\verb|WHERE | \{}\\
			\scriptsize{~~\verb|?X rdf:type ub:Student.|}\\
			\scriptsize{~~\verb|?Y rdf:type ub:Department.|}\\
			\scriptsize{~~\verb|?X ub:memberOf ?Y.|}\\
			\scriptsize{~~\verb|?Y ub:subOrganizationOf <http://www.University0.edu>.|}\\
			\scriptsize{~~\verb|?X ub:emailAddress ?Z| \}}\\
			\includegraphics[scale=0.67]{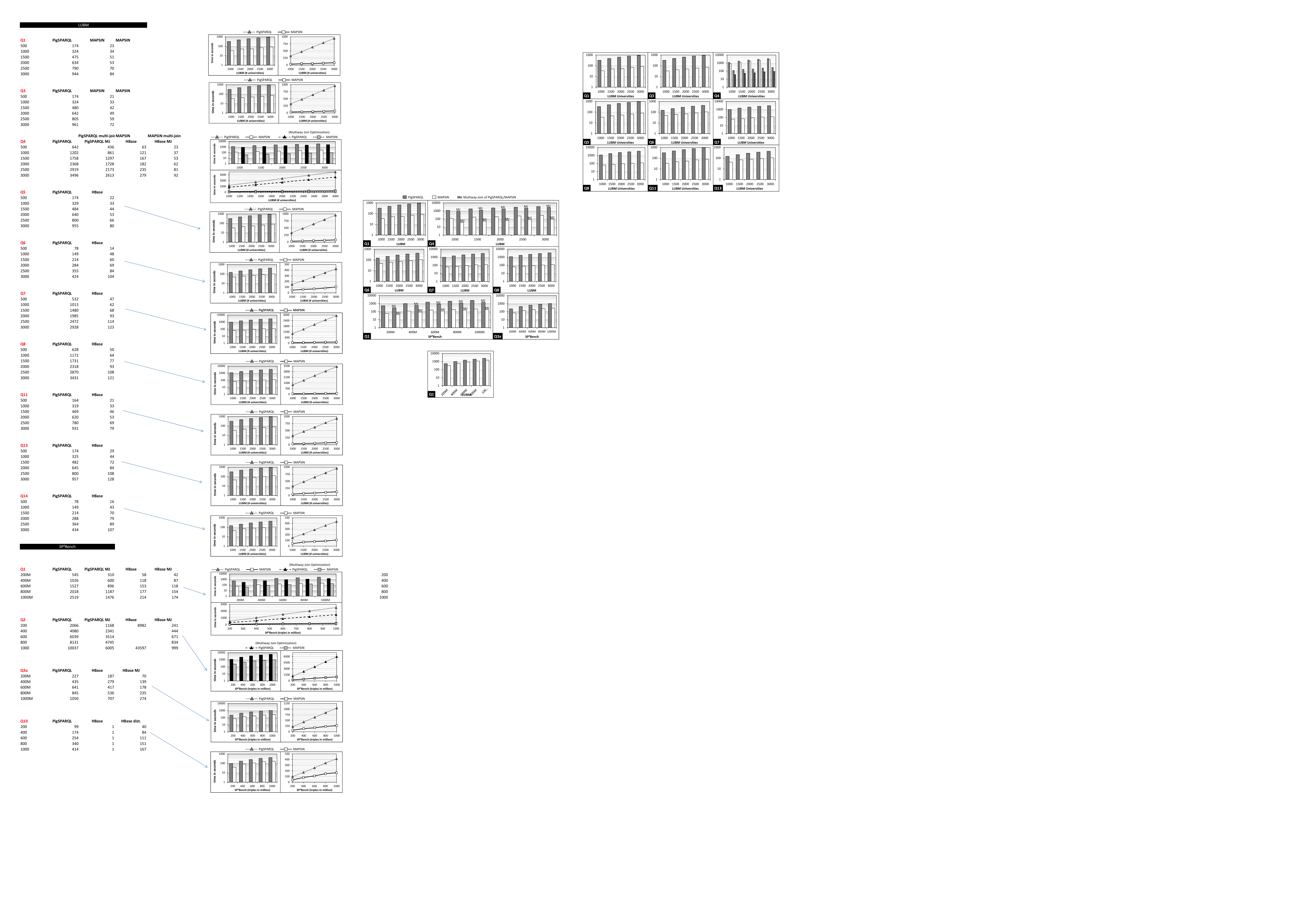}\\
			\hline	
		\end{tabularx}
	\caption{Performance comparison for LUBM Q8}
	\label{fig:eval_q8}
\end{figure}

LUBM queries Q4 (5 triple patterns), Q7 (4 triple patterns), Q8 (5 triple patterns) and SP$^2$Bench queries Q1 (3 triple patterns), Q2 (9 triple patterns) demonstrate the more general case with a sequence of cascaded joins (cf. Figure~\ref{fig:eval_q8}). In these cases, MAPSIN joins perform even up to 28 times faster than PigSPARQL for LUBM queries and up to 12 times faster for SP$^2$Bench queries.

\begin{table*}[htbp]
	\centering
	\caption{Query execution times for PigSPARQL (P) and MAPSIN (M) in seconds}
	\label{tab:ExecutionTimes}
		\begin{tabularx}{\textwidth}{|X|c|c|c|c|c|c|c|c|c|c|c|c|}
			\hline			
			\textbf{LUBM} & \multicolumn{2}{c|}{\textbf{1000}} & \multicolumn{2}{c|}{\textbf{1500}} & \multicolumn{2}{c|}{\textbf{2000}} & \multicolumn{2}{c|}{\textbf{2500}} & \multicolumn{2}{c|}{\textbf{3000}}\\
			\hline
			              &   ~~~~P~~~~ & ~~~~M~~~~  &  ~~~~P~~~~ & ~~~~M~~~~  &  ~~~~P~~~~ & ~~~M~~~~  &  ~~~~P~~~~ & ~~~~M~~~~  &  ~~~~P~~~~ & ~~~~M~~~~\\
			\hline
			Q1            &   324 & 34     &  475 & 51      &  634 & 53     &  790 & 70     &  944 & 84\\
			\hline
			Q3            &   324 & 33     &  480 & 42      &  642 & 49     &  805 & 59     &  961 & 72\\
			\hline
			Q4            &   1202 & 121   &  1758 & 167    &  2368 & 182   &  2919 & 235   &  3496 & 279\\
			\hline
			Q4 multijoin  &   861 & 37     &  1297 & 53     &  1728 & 62    &  2173 & 81    &  2613 & 92\\
			\hline
			Q5            &   329 & 33     &  484 & 44      &  640 & 53     &  800 & 66     &  955 & 80\\
			\hline
			Q6            &   149 & 48     &  214 & 60      &  284 & 69     &  355 & 84     &  424 & 104\\
			\hline
			Q7            &   1013 & 62    &  1480 & 68     &  1985 & 93    &  2472 & 114   &  2928 & 123\\
			\hline
			Q8            &   1172 & 64    &  1731 & 77     &  2318 & 33    &  2870 & 108   &  3431 & 121\\
			\hline
			Q11           &   319 & 33     &  469 & 46      &  620 & 53     &  780 & 69     &  931 & 79\\
			\hline
			Q13           &   325 & 44     &  482 & 72      &  645 & 84     &  800 & 108    &  957 & 128\\
			\hline
			Q14           &   149 & 43     &  214 & 70      &  288 & 79     &  364 & 89     &  434 & 107\\
			\hline
			\hline
			\textbf{SP$^2$Bench} & \multicolumn{2}{c|}{\textbf{200M}} & \multicolumn{2}{c|}{\textbf{400M}} & \multicolumn{2}{c|}{\textbf{600M}} & \multicolumn{2}{c|}{\textbf{800M}} & \multicolumn{2}{c|}{\textbf{1000M}}\\
			\hline
			              &   ~~~~P~~~~ & ~~~~M~~~~  &  ~~~~P~~~~ & ~~~~M~~~~  &  ~~~~P~~~~ & ~~~M~~~~  &  ~~~~P~~~~ & ~~~~M~~~~  &  ~~~~P~~~~ & ~~~~M~~~~\\
			\hline
			Q1            &   545 & 58     &  1026 & 118    &  1527 & 153   &  2018 & 177   &  2519 & 214\\
			\hline
			Q1 multijoin  &   310 & 42     &  600 & 87      &  896 & 118    &  1187 & 154   &  1476 & 174\\
			\hline
			Q2 multijoin  &   1168 & 241   &  2341 & 444    &  3514 & 671   &  4745 & 834   &  6005 & 999\\
			\hline
			Q3a           &   227 & 70     &  435 & 139     &  641 & 178    &  845 & 235    &  1050 & 274\\
			\hline
			Q10           &   99 & 40      &  174 & 84      &  254 & 111    &  340 & 151    &  414 & 167\\
			\hline
		\end{tabularx}
\end{table*}

\begin{figure}[h]
	\centering
		\begin{tabularx}{\columnwidth}{|X|}
			\hline
			\rowcolor{gray} \textbf{LUBM Q4.} \textit{Return names, emails and phone numbers of all professors working for a given department}\\
			\hline
			\scriptsize{\verb|SELECT ?X, ?Y1, ?Y2, ?Y3|}\\
			\scriptsize{\verb|WHERE | \{}\\
			\scriptsize{~~\verb|?X rdf:type ub:Professor.|}\\
			\scriptsize{~~\verb|?X ub:worksFor <http://www.Department0.University0.edu>.|}\\
			\scriptsize{~~\verb|?X ub:name ?Y1.|}\\
			\scriptsize{~~\verb|?X ub:emailAddress ?Y2.|}\\
			\scriptsize{~~\verb|?X ub:telephone ?Y3| \}}\\
			\includegraphics[scale=0.67]{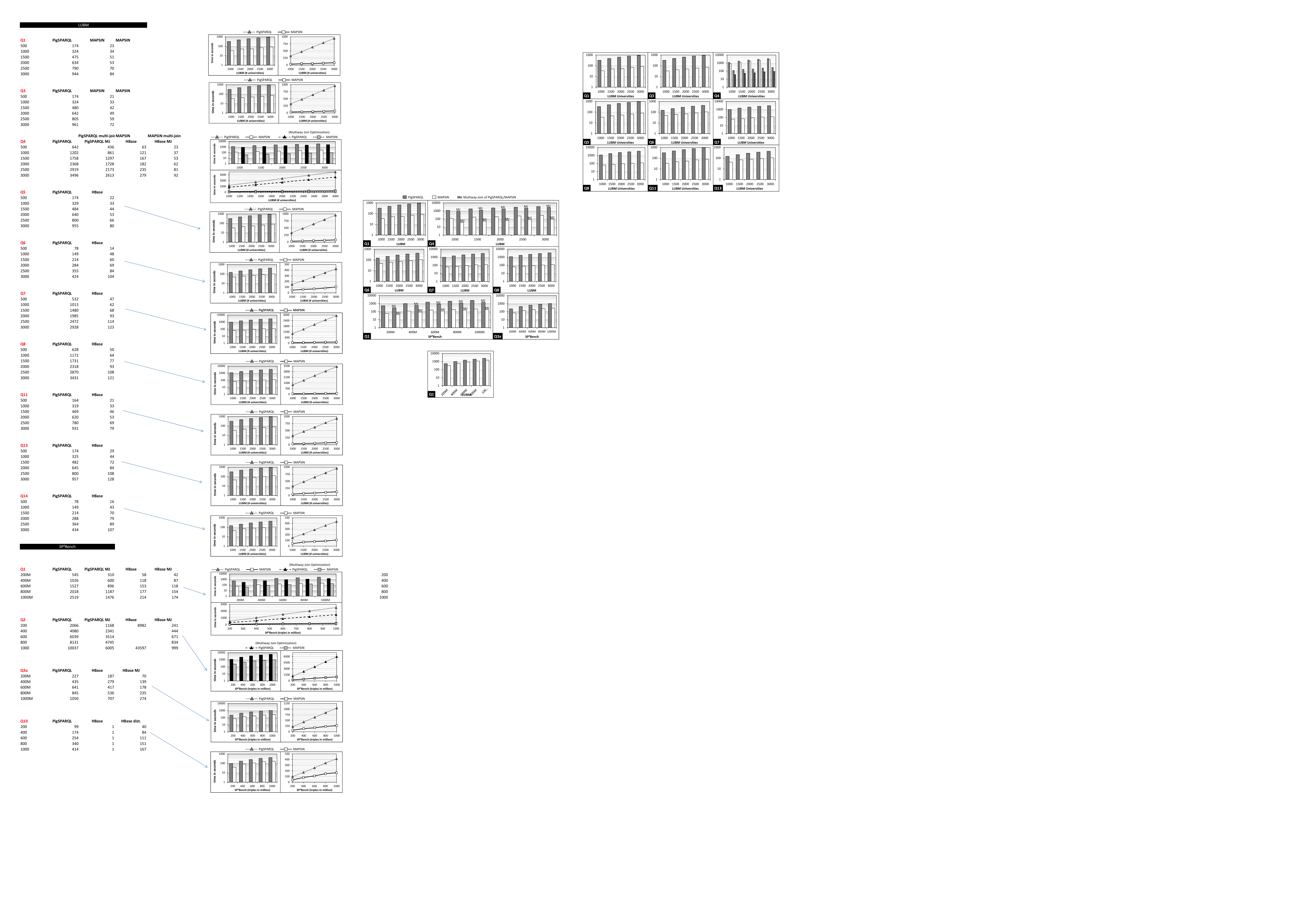}\\
			\hline	
		\end{tabularx}
	\caption{Performance comparison for LUBM Q4}
	\label{fig:eval_q4}
\end{figure}

Especially interesting are queries Q4 of LUBM and Q1, Q2 of SP$^2$Bench since these queries support the multiway join optimization outlined in Section \ref{subsec:MAPSIN_MULTI} as all triple patterns share the same join variable. This kind of optimization is also supported by PigSPARQL such that both approaches can compute the query results with a single multiway join (cf. Figure~\ref{fig:eval_q4}). The MAPSIN multiway join optimization improves the basic MAPSIN join execution time by a factor of 1.4 (SP$^2$Bench Q1) to 3.3 (LUBM Q4), independently of the data size. For the LUBM queries, the MAPSIN multiway join optimization performs 19 to 28 times faster than the reduce-side based multiway join implementation of PigSPARQL. For the more complex SP$^2$Bench queries, the performance improvements degrade to a factor of approximately 8.5.

\begin{figure}[h]
	\centering
		\begin{tabularx}{\columnwidth}{|X|}
			\hline
			\rowcolor{gray} \textbf{LUBM Q6.} \textit{Return all students}\\
			\hline
			\scriptsize{\verb|SELECT ?X|}\\
			\scriptsize{\verb|WHERE | \{ \verb|?X rdf:type ub:Student| \}}\\
			\includegraphics[scale=0.67]{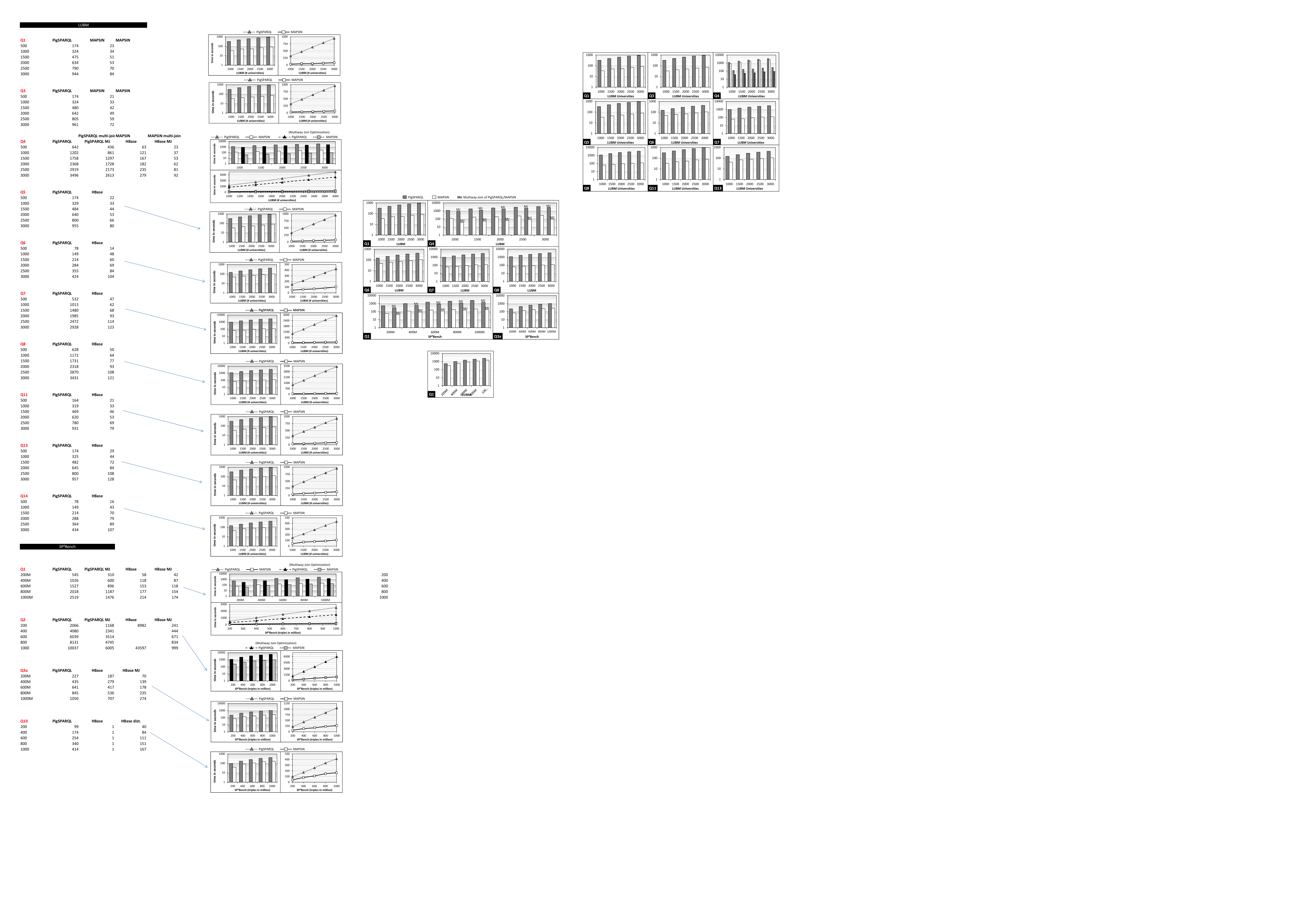}\\
			\hline	
		\end{tabularx}
	\caption{Performance comparison for LUBM Q6}
	\label{fig:eval_q6}
\end{figure}

The remaining queries (LUBM Q6, Q14 and SP$^2$Bench Q10) consist of only one single triple pattern (cf. Figure~\ref{fig:eval_q6}). Consequently they do not contain a join processing step and illustrate primarily the advantages of the distributed HBase table scan compared to the HDFS storage access of PigSPARQL. Improvements are still present but less significant, resulting in an up to 5 times faster query execution.

Overall, the MAPSIN join approach clearly outperforms the reduce-side joins of PigSPARQL. Both approaches reveal a linear scaling behavior with the input size but the slope of the MAPSIN join is much smaller. Especially for LUBM queries, MAPSIN joins outperform reduce-side joins by an order of magnitude as these queries are generally rather selective. Moreover, the application of our multiway join optimization results in a further significant improvement of the total query execution times.

\newpage
\section{Related Work}
\label{sec-relatedwork}
Single machine RDF engines like \emph{Sesame}~\cite{sesame2002} and \emph{Jena}~\cite{jena2003} are widely-used since they are user-friendly and perform well for small and medium sized RDF datasets. \emph{RDF-3X}~\cite{rdf-3x} is considered one of the fastest single machine RDF engines in terms of query performance that vastly outperforms previous single machine systems but performance degrades for queries with unbound objects and low selectivity factor~\cite{DBLP:journals/tkde/HusainMMKT11}. Furthermore, as the amount of RDF data continues to grow, it will become more and more difficult to store entire datasets on a single machine due to the limited scaling capabilities~\cite{DBLP:journals/pvldb/HuangAR11}.
\emph{OWLIM}~\cite{owlim} is a fast and scalable RDF database management system that offers both a single and a cluster edition but it requires high end and thus costly hardware~\cite{DBLP:journals/tkde/HusainMMKT11}.

There is also a large body of work dealing with join processing in a MapReduce environment considering various aspects and application fields~\cite{DBLP:journals/tkde/AfratiU11,DBLP:conf/sigmod/BlanasPERST10,DBLP:journals/tkde/JiangTC11,DBLP:journals/sigmod/LeeLCCM11,DBLP:conf/sigmod/OkcanR11,DBLP:conf/esws/Przyjaciel-ZablockiSHL11,DBLP:conf/sigmod/YangDHP07}.
In Section~\ref{subsec:MapReduce} we briefly outlined the advantages and drawbacks of the general-purpose reduce-side and map-side (merge) join approaches in MapReduce. Though map-side joins are generally more efficient, they are hard to cascade due to the strict preconditions. In our approach, we leverage HBase to overcome this shortcoming without the use of auxiliary shuffle and reduce phases, making MAPSIN joins easily cascadable.
In addition to these general-purpose approaches there are several proposals focusing on certain join types or optimizations of existing join techniques for particular application fields.
In~\cite{DBLP:conf/sigmod/OkcanR11} the authors discussed how to process arbitrary joins (theta joins) using MapReduce and provide algorithms to minimize total job completion time, whereas  \cite{DBLP:journals/tkde/AfratiU11} focuses on optimizing multiway joins. However, in contrast to our MAPSIN join, both approaches process the join in the reduce phase including a costly data shuffle phase.
\emph{Map-Reduce-Merge}~\cite{DBLP:conf/sigmod/YangDHP07} describes a modified MapReduce workflow by adding a merge phase after the reduce phase, whereas \emph{Map-Join-Reduce}~\cite{DBLP:journals/tkde/JiangTC11} proposes a join phase in between the map and reduce phase. Both techniques attempt to improve the support for joins in MapReduce but require profound modifications to the MapReduce framework. Our approach requires no changes to Hadoop and HBase at all, thus making it possible to run on any Hadoop cluster out of the box, e.g. Amazon's Elastic Compute Cloud (EC2)\footnote{\url{http://aws.amazon.com/ec2/}}.
In~\cite{DBLP:journals/pvldb/DittrichQJKSS10} the authors propose a new index and join technique called \emph{Trojan Index} and \emph{Trojan Join} that do not require modifications to the MapReduce framework. These techniques reduce the amount of data that is shuffled during join execution at the cost of an additional co-partitioning and indexing phase at load time. However, in contrast to our approach, shuffle and reduce phases cannot be completely avoided. 

\emph{HadoopDB}~\cite{DBLP:journals/pvldb/AbouzeidBARS09} is a hybrid of MapReduce and DBMS where MapReduce is the communication layer above multiple single node DBMS. The authors in \cite{DBLP:journals/pvldb/HuangAR11} adopt this hybrid approach for the semantic web using RDF-3X. However, the initial graph partitioning is done on a single machine and has to be repeated if the dataset is updated or the number of machines in the cluster change. As we use HBase as underlying storage layer, additional machines can be plugged in seamlessly and updates are possible without having to reload the entire dataset.

\emph{HadoopRDF}~\cite{DBLP:journals/tkde/HusainMMKT11} and \emph{SHARD}~\cite{shard} are MapReduce based RDF engines that store data directly in HDFS and do not require any changes to the Hadoop framework. These systems are able to rebalance automatically if cluster size changes but join processing is also done in the reduce phase. Our MAPSIN join does not use any shuffle or reduce phase at all even in consecutive iterations. As we only request the data from HBase that is really needed, we reduce data transfer cost to a minimum.

\section{Conclusion}
\label{sec-conclusion}
In this paper we introduced the Map-Side Index Nested Loop join (MAPSIN join) which combines the advantages of NoSQL storage systems like HBase with the well-known and approved distributed processing facilities of MapReduce. In general, map-side joins are more efficient than reduce-side joins in MapReduce as there is no expensive data shuffle phase involved. However, current map-side join approaches have some strict preconditions what makes them hard to apply in general, especially in a sequence of joins. The combination of HBase and MapReduce allows us to cascade a sequence of MAPSIN joins without having to reorder or sort the output for the next iteration. Furthermore, with the multiway join optimization we can reduce the number of MapReduce iterations and HBase requests. Using an index to selectively request only those data that is really needed also saves network bandwidth, making parallel query execution more efficient. The evaluation with the LUBM and SP$^2$Bench benchmarks demonstrate the advantages of our approach compared to the commonly used reduce-side join approach in MapReduce. For most of the benchmark queries, MAPSIN join based SPARQL query execution outperforms reduce-side join based execution of PigSPARQL (using Apache Pig) by an order of magnitude while scaling very smoothly with the input size. Lastly, our approach does not require any changes of the Hadoop platform. Consequently, SPARQL queries can be run directly on an instance of Amazon's Elastic Compute Cloud (EC2) whenever the own capacities are exceeded.

In our future work, we will investigate alternatives and improvements of the RDF storage schema for HBase and incorporate MAPSIN joins into PigSPARQL in a hybrid fashion such that the actual join method is dynamically selected based on pattern selectivity and statistics gathered at data loading time.

\bibliographystyle{abbrv}
\bibliography{MAPSIN}

\begin{thebibliography}{10}

\bibitem{rdfprimer}
{RDF Primer. W3C Recommendation}.
\newblock \url{http://www.w3.org/TR/rdf-primer/}, 2004.

\bibitem{sparql}
{SPARQL Query Language for RDF. W3C Recom.}
\newblock \url{http://www.w3.org/TR/rdf-sparql-query/}, 2008.

\bibitem{DBLP:journals/pvldb/AbouzeidBARS09}
A.~Abouzeid, K.~Bajda-Pawlikowski, D.~J. Abadi, A.~Rasin, and A.~Silberschatz.
\newblock {HadoopDB: An Architectural Hybrid of MapReduce and DBMS Technologies
  for Analytical Workloads}.
\newblock {\em PVLDB}, 2(1):922--933, 2009.

\bibitem{DBLP:journals/tkde/AfratiU11}
F.~N. Afrati and J.~D. Ullman.
\newblock {Optimizing Multiway Joins in a Map-Reduce Environment}.
\newblock {\em IEEE Trans. Knowl. Data Eng.}, 23(9):1282--1298, 2011.

\bibitem{bernerslee2001semantic}
T.~Berners-Lee, J.~Hendler, and O.~Lassila.
\newblock {The Semantic Web}.
\newblock {\em Scientific American}, 284(5):34--43, 2001.

\bibitem{DBLP:conf/sigmod/BlanasPERST10}
S.~Blanas, J.~M. Patel, V.~Ercegovac, J.~Rao, E.~J. Shekita, and Y.~Tian.
\newblock {A Comparison of Join Algorithms for Log Processing in MapReduce}.
\newblock In {\em SIGMOD Conference}, pages 975--986, 2010.

\bibitem{sesame2002}
J.~Broekstra, A.~Kampman, and F.~van Harmelen.
\newblock {Sesame: A Generic Architecture for Storing and Querying RDF and RDF
  Schema}.
\newblock In {\em International Semantic Web Conference (ISWC)}, pages 54--68,
  2002.

\bibitem{chang2008bigtable}
F.~Chang, J.~Dean, S.~Ghemawat, W.~Hsieh, D.~Wallach, M.~Burrows, T.~Chandra,
  A.~Fikes, and R.~Gruber.
\newblock {Bigtable: A Distributed Storage System for Structured Data}.
\newblock {\em ACM Transactions on Computer Systems (TOCS)}, 26(2):4, 2008.

\bibitem{dean2008mapreduce}
J.~Dean and S.~Ghemawat.
\newblock {MapReduce: Simplified Data Processing on Large Clusters}.
\newblock {\em Communications of the ACM}, 51(1):107--113, 2008.

\bibitem{DBLP:journals/pvldb/DittrichQJKSS10}
J.~Dittrich, J.-A. Quian{\'e}-Ruiz, A.~Jindal, Y.~Kargin, V.~Setty, and
  J.~Schad.
\newblock {Hadoop++: Making a Yellow Elephant Run Like a Cheetah (Without It
  Even Noticing)}.
\newblock {\em PVLDB}, 3(1):518--529, 2010.

\bibitem{franke2011}
C.~Franke, S.~Morin, A.~Chebotko, J.~Abraham, and P.~Brazier.
\newblock {Distributed Semantic Web Data Management in HBase and MySQL
  Cluster}.
\newblock In {\em IEEE International Conference on Cloud Computing (CLOUD)},
  pages 105 --112, 2011.

\bibitem{gates2009}
A.~F. Gates, O.~Natkovich, S.~Chopra, P.~Kamath, S.~M. Narayanamurthy,
  C.~Olston, B.~Reed, S.~Srinivasan, and U.~Srivastava.
\newblock {Building a High-Level Dataflow System on top of Map-Reduce: The Pig
  Experience}.
\newblock {\em PVLDB}, 2(2):1414--1425, 2009.

\bibitem{george2011hbase}
L.~George.
\newblock {\em {HBase - The Definitive Guide: Random Access to Your Planet-Size
  Data}}.
\newblock O'Reilly, 2011.

\bibitem{ghemawat2003google}
S.~Ghemawat, H.~Gobioff, and S.~Leung.
\newblock {The Google File System}.
\newblock In {\em ACM SIGOPS Operating Systems Review}, volume~37, pages
  29--43. ACM, 2003.

\bibitem{guo2005lubm}
Y.~Guo, Z.~Pan, and J.~Heflin.
\newblock {LUBM: A Benchmark for OWL Knowledge Base Systems}.
\newblock {\em Web Semantics: Science, Services and Agents on the World Wide
  Web}, 3(2):158--182, 2005.

\bibitem{hewitt2010cassandra}
E.~Hewitt.
\newblock {\em {Cassandra - The Definitive Guide: Distributed Data at Web
  Scale}}.
\newblock Definitive Guide Series. O'Reilly, 2010.

\bibitem{DBLP:journals/pvldb/HuangAR11}
J.~Huang, D.~J. Abadi, and K.~Ren.
\newblock {Scalable SPARQL Querying of Large RDF Graphs}.
\newblock {\em PVLDB}, 4(11):1123--1134, 2011.

\bibitem{DBLP:journals/tkde/HusainMMKT11}
M.~F. Husain, J.~P. McGlothlin, M.~M. Masud, L.~R. Khan, and B.~M.
  Thuraisingham.
\newblock {Heuristics-Based Query Processing for Large RDF Graphs Using Cloud
  Computing}.
\newblock {\em IEEE Trans. Knowl. Data Eng.}, 23(9):1312--1327, 2011.

\bibitem{DBLP:journals/tkde/JiangTC11}
D.~Jiang, A.~K.~H. Tung, and G.~Chen.
\newblock {Map-Join-Reduce: Toward Scalable and Efficient Data Analysis on
  Large Clusters}.
\newblock {\em IEEE Trans. Knowl. Data Eng.}, 23(9):1299--1311, 2011.

\bibitem{owlim}
A.~Kiryakov, D.~Ognyanov, and D.~Manov.
\newblock {OWLIM - A Pragmatic Semantic Repository for OWL}.
\newblock In {\em WISE Workshops}, pages 182--192, 2005.

\bibitem{DBLP:journals/sigmod/LeeLCCM11}
K.-H. Lee, Y.-J. Lee, H.~Choi, Y.~D. Chung, and B.~Moon.
\newblock {Parallel Data Processing with MapReduce: A Survey}.
\newblock {\em SIGMOD Record}, 40(4):11--20, 2011.

\bibitem{2010Lin}
J.~Lin and C.~Dyer.
\newblock {\em {Data-Intensive Text Processing with MapReduce}}.
\newblock Synthesis Lectures on Human Language Technologies. Morgan {\&}
  Claypool Publishers, 2010.

\bibitem{rdf-3x}
T.~Neumann and G.~Weikum.
\newblock {RDF-3X: a RISC-style engine for RDF}.
\newblock {\em PVLDB}, 1(1):647--659, 2008.

\bibitem{DBLP:conf/sigmod/OkcanR11}
A.~Okcan and M.~Riedewald.
\newblock {Processing Theta-Joins using MapReduce}.
\newblock In {\em SIGMOD Conference}, pages 949--960, 2011.

\bibitem{perez2009semantics}
J.~P{\'e}rez, M.~Arenas, and C.~Gutierrez.
\newblock {Semantics and Complexity of SPARQL}.
\newblock {\em ACM Transactions on Database Systems (TODS)}, 34(3):16, 2009.

\bibitem{DBLP:conf/esws/Przyjaciel-ZablockiSHL11}
M.~Przyjaciel-Zablocki, A.~Sch{\"a}tzle, T.~Hornung, and G.~Lausen.
\newblock {RDFPath: Path Query Processing on Large RDF Graphs with MapReduce}.
\newblock In {\em ESWC Workshops}, pages 50--64, 2011.

\bibitem{shard}
K.~Rohloff and R.~E. Schantz.
\newblock {High-Performance, Massively Scalable Distributed Systems using the
  MapReduce Software Framework: The SHARD Triple-Store}.
\newblock In {\em Programming Support Innovations for Emerging Distributed
  Applications (PSI EtA)}, pages 4:1--4:5, 2010.

\bibitem{PigSPARQL}
A.~Sch\"{a}tzle, M.~Przyjaciel-Zablocki, and G.~Lausen.
\newblock {PigSPARQL: Mapping SPARQL to Pig Latin}.
\newblock In {\em Proceedings of the International Workshop on Semantic Web
  Information Management (SWIM)}, pages 4:1--4:8, 2011.

\bibitem{schmidt2009sp}
M.~Schmidt, T.~Hornung, G.~Lausen, and C.~Pinkel.
\newblock {SP2Bench: A SPARQL Performance Benchmark}.
\newblock In {\em IEEE 25th International Conference on Data Engineering
  (ICDE)}, pages 222--233, 2009.

\bibitem{stocker2008}
M.~Stocker, A.~Seaborne, A.~Bernstein, C.~Kiefer, and D.~Reynolds.
\newblock {SPARQL Basic Graph Pattern Optimization Using Selectivity
  Estimation}.
\newblock In {\em Proceedings of the 17th international conference on World
  Wide Web (WWW)}, pages 595--604, 2008.

\bibitem{sun2010}
J.~Sun and Q.~Jin.
\newblock {Scalable RDF Store Based on HBase and MapReduce}.
\newblock In {\em 3rd International Conference on Advanced Computer Theory and
  Engineering (ICACTE)}, volume~1, pages 633--636, 2010.

\bibitem{urbani2010owl}
J.~Urbani, S.~Kotoulas, J.~Maassen, F.~van Harmelen, and H.~Bal.
\newblock {OWL Reasoning with WebPIE: Calculating the Closure of 100 Billion
  Triples}.
\newblock In {\em ESWC}, pages 213--227, 2010.

\bibitem{weiss2008}
C.~Weiss, P.~Karras, and A.~Bernstein.
\newblock {Hexastore: Sextuple Indexing for Semantic Web Data Management}.
\newblock {\em PVLDB}, 1(1):1008--1019, 2008.

\bibitem{hadoopGuideWhite}
T.~White.
\newblock {\em {Hadoop - The Definitive Guide: Storage and Analysis at Internet
  Scale (2. ed.)}}.
\newblock O'Reilly, 2011.

\bibitem{jena2003}
K.~Wilkinson, C.~Sayers, H.~A. Kuno, and D.~Reynolds.
\newblock {Efficient RDF Storage and Retrieval in Jena2}.
\newblock In {\em SWDB}, pages 131--150, 2003.

\bibitem{DBLP:conf/sigmod/YangDHP07}
H.-C. Yang, A.~Dasdan, R.-L. Hsiao, and D.~S.~P. Jr.
\newblock {Map-Reduce-Merge: Simplified Relational Data Processing on Large
  Clusters}.
\newblock In {\em SIGMOD Conference}, pages 1029--1040, 2007.

\end{thebibliography}

\clearpage
\appendix
\section{LUBM}
\label{appendix-lubm}
\begin{table}[h!]
	\centering
		\begin{tabularx}{\columnwidth}{|X|}
			\hline
			\rowcolor{gray} \textbf{Q1.} \textit{Return all graduate students that attend a given course from a given university department}\\
			\hline
			\scriptsize{\verb|SELECT ?X|}\\
			\scriptsize{\verb|WHERE | \{}\\
			\scriptsize{~~\verb|?X rdf:type ub:GraduateStudent.|}\\
			\scriptsize{~~\verb|?X ub:takesCourse|}\\
			\scriptsize{~~~~~~\verb|<http://www.Department0.University0.edu/GraduateCourse0>|}\\
			\scriptsize{\}}\\		
			\includegraphics[scale=0.67]{evaluation/lubm_q1}\\		
			\hline	
		\end{tabularx}
		
		\begin{tabularx}{\columnwidth}{X}
		\verb| |\\
		\end{tabularx}
		
		\begin{tabularx}{\columnwidth}{|X|}
			\hline
			\rowcolor{gray} \textbf{Q3.} \textit{Return all publications of a given assistant professor from a given university department}\\
			\hline
			\scriptsize{\verb|SELECT ?X|}\\
			\scriptsize{\verb|WHERE | \{}\\
			\scriptsize{~~\verb|?X rdf:type ub:Publication.|}\\
			\scriptsize{~~\verb|?X ub:publicationAuthor|}\\
			\scriptsize{~~~~~~\verb|<http://www.Department0.University0.edu/AssistantProfessor0>|}\\
			\scriptsize{\}}\\
			\includegraphics[scale=0.67]{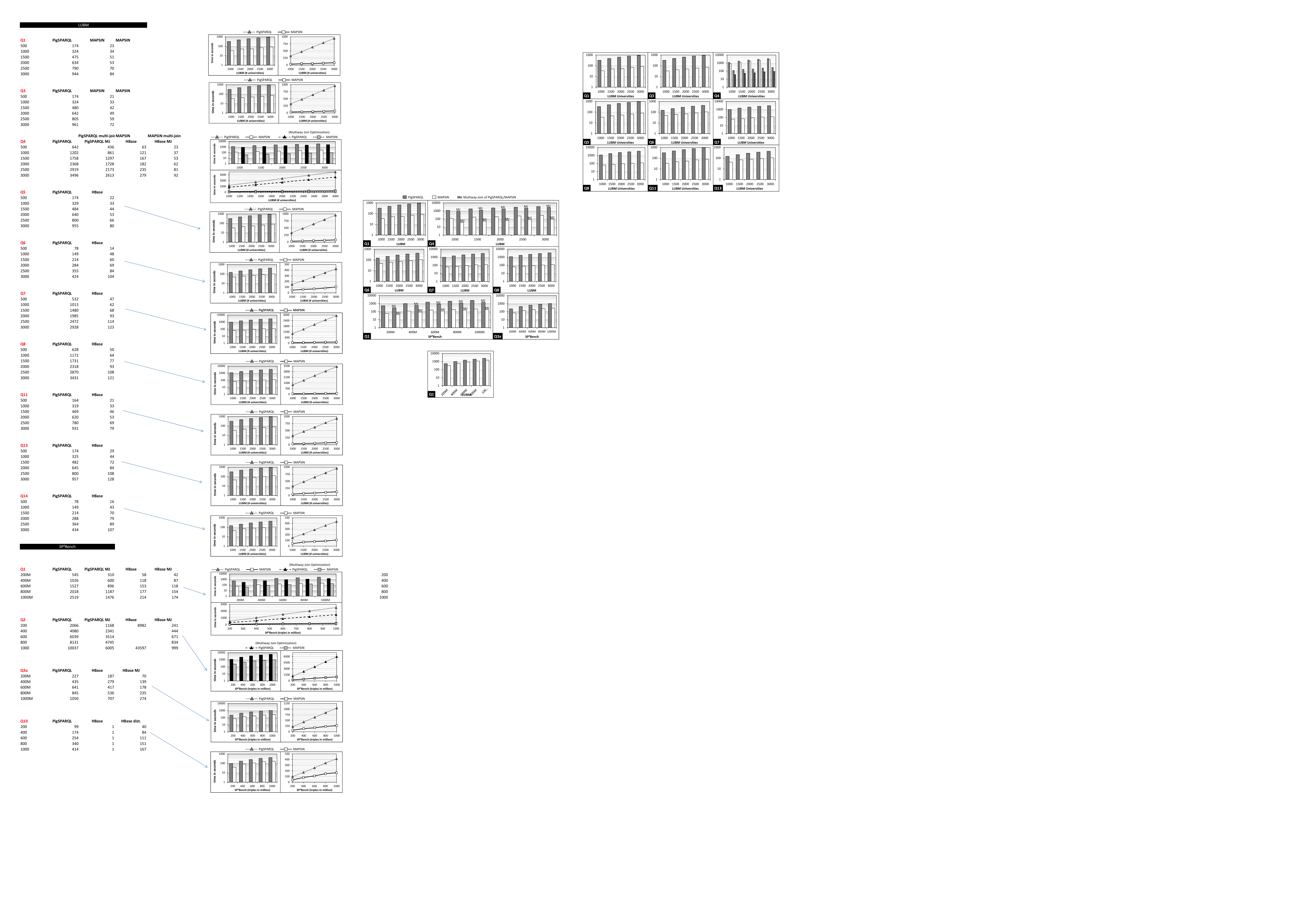}\\
			\hline
		\end{tabularx}
		
		\begin{tabularx}{\columnwidth}{X}
		\verb| |\\
		\end{tabularx}
		
		\begin{tabularx}{\columnwidth}{|X|}
			\hline
			\rowcolor{gray} \textbf{Q4.} \textit{Return all professors with name, email address and telephone number working for a given university department}\\
			\hline
			\scriptsize{\verb|SELECT ?X ?Y1 ?Y2 ?Y3|}\\
			\scriptsize{\verb|WHERE | \{}\\
			\scriptsize{~~\verb|?X rdf:type ub:Professor.|}\\
			\scriptsize{~~\verb|?X ub:worksFor <http://www.Department0.University0.edu>.|}\\
			\scriptsize{~~\verb|?X ub:name ?Y1.|}\\
			\scriptsize{~~\verb|?X ub:emailAddress ?Y2.|}\\
			\scriptsize{~~\verb|?X ub:telephone ?Y3|}\\
			\scriptsize{\}}\\
			\includegraphics[scale=0.67]{evaluation/lubm_q4}\\
			\hline			
		\end{tabularx}		
\end{table}

\begin{table}[h!]
	\centering
		\begin{tabularx}{\columnwidth}{|X|}
			\hline
			\rowcolor{gray} \textbf{Q5.} \textit{Return all persons that are members of a given university department}\\
			\hline
			\scriptsize{\verb|SELECT ?X|}\\
			\scriptsize{\verb|WHERE | \{}\\
			\scriptsize{~~\verb|?X rdf:type ub:Person.|}\\
			\scriptsize{~~\verb|?X ub:memberOf <http://www.Department0.University0.edu>|}\\
			\scriptsize{\}}\\
			\includegraphics[scale=0.67]{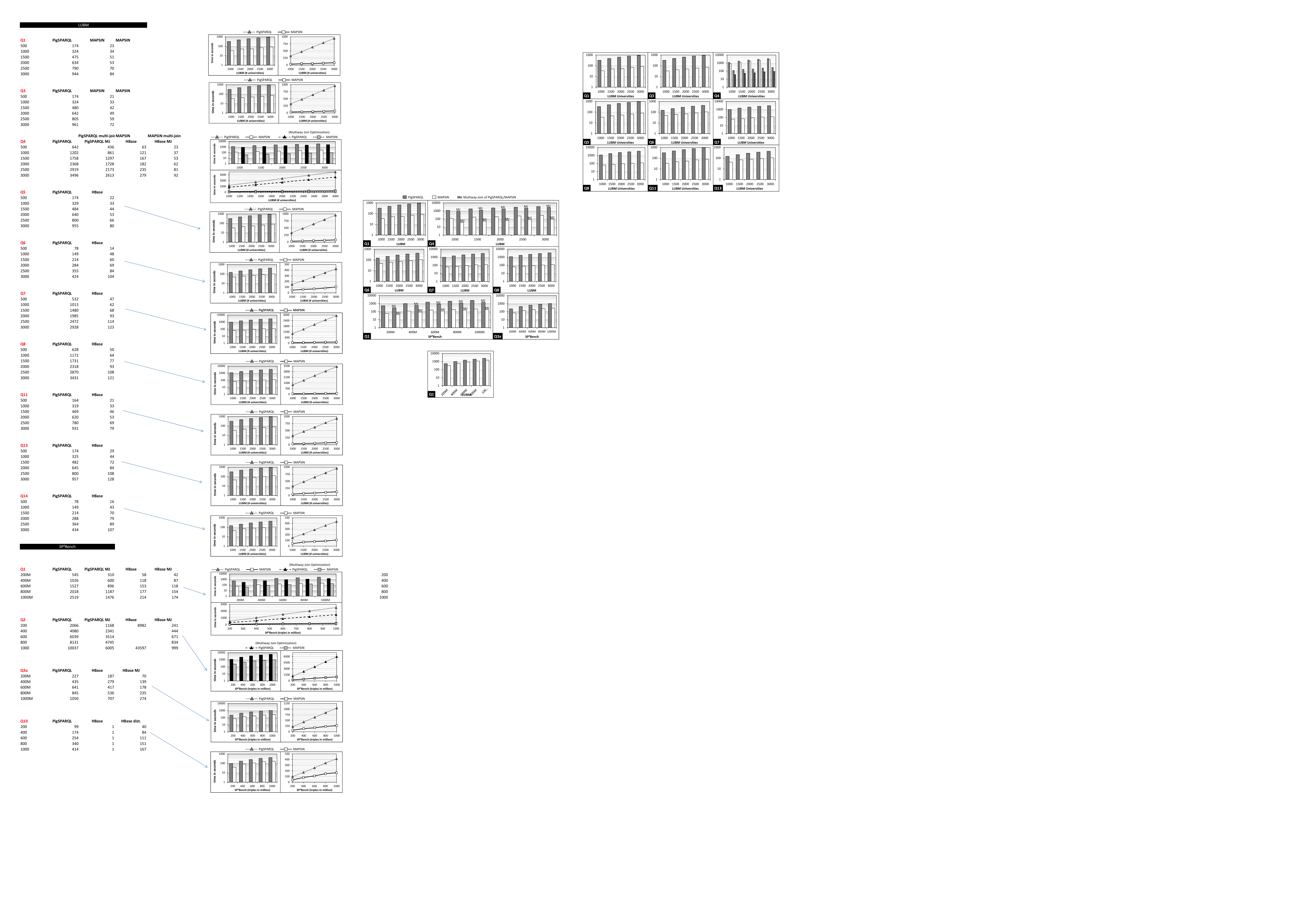} \\
			\hline				
		\end{tabularx}
		
		\begin{tabularx}{\columnwidth}{X}
		\verb| |\\
		\end{tabularx}		
		
		\begin{tabularx}{\columnwidth}{|X|}
			\hline
			\rowcolor{gray} \textbf{Q6.} \textit{Return all students}\\
			\hline
			\scriptsize{\verb|SELECT ?X|}\\
			\scriptsize{\verb|WHERE | \{}\\
			\scriptsize{~~\verb|?X rdf:type ub:Student|}\\
			\scriptsize{\}}\\
			\includegraphics[scale=0.67]{evaluation/lubm_q6}\\
			\hline			
		\end{tabularx}
		
		\begin{tabularx}{\columnwidth}{X}
		\verb| |\\
		\end{tabularx}	
		
		\begin{tabularx}{\columnwidth}{|X|}
			\hline
			\rowcolor{gray} \textbf{Q7.} \textit{Return all students that attend any course of a given associate professor from a given university department}\\
			\hline
			\scriptsize{\verb|SELECT ?X ?Y|}\\
			\scriptsize{\verb|WHERE | \{}\\
			\scriptsize{~~\verb|?X rdf:type ub:Student.|}\\
			\scriptsize{~~\verb|?Y rdf:type ub:Course.|}\\
			\scriptsize{~~\verb|?X ub:takesCourse ?Y.|}\\
			\scriptsize{~~\verb|<http://www.Department0.University0.edu/AssociateProfessor0>|}\\
			\scriptsize{~~~~~~\verb|ub:teacherOf ?Y|}\\
			\scriptsize{\}}\\
			\includegraphics[scale=0.67]{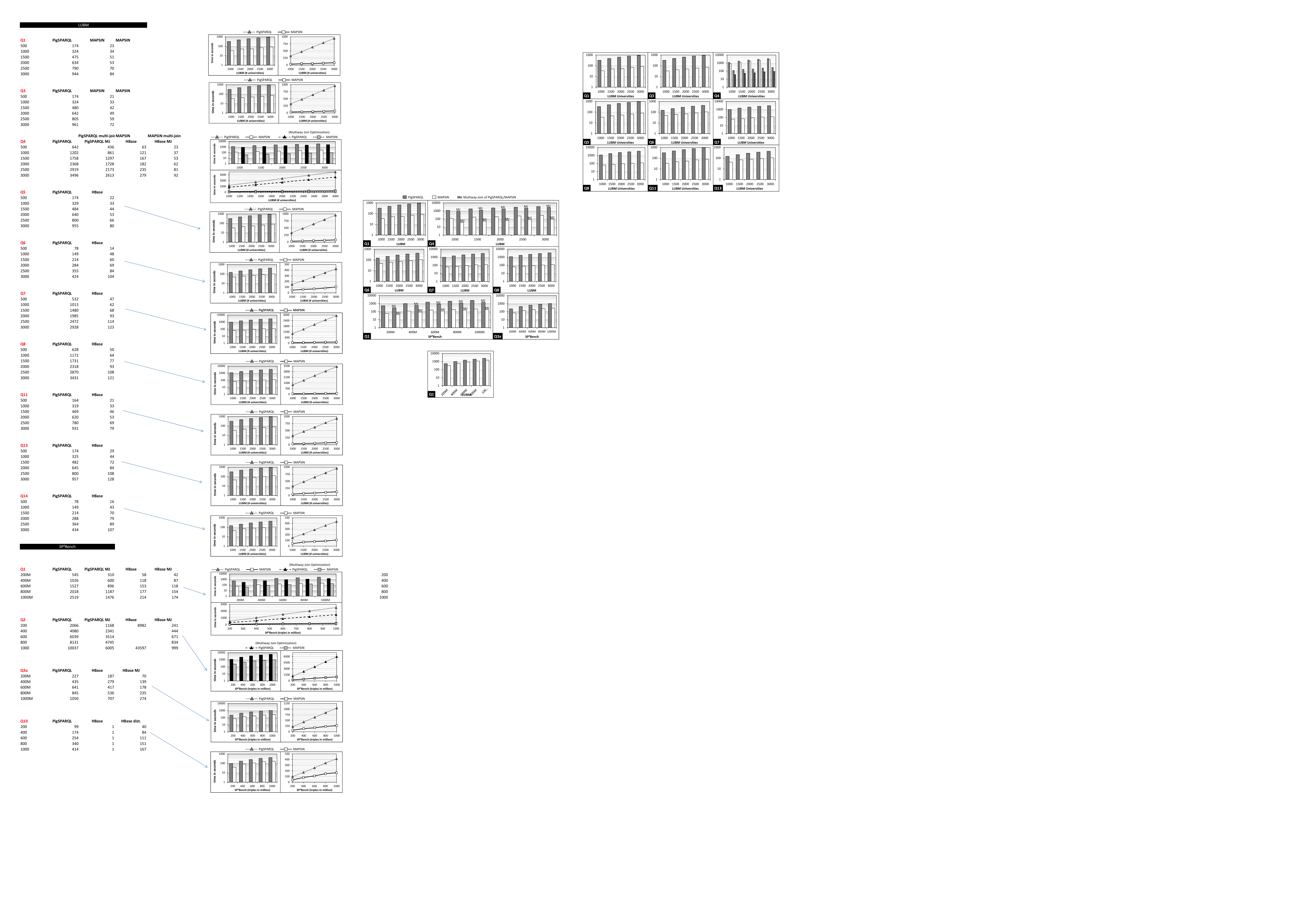}\\
			\hline				
		\end{tabularx}
		
		\begin{tabularx}{\columnwidth}{X}
		\verb| |\\
		\verb| |\\
		\verb| |\\
		\verb| |\\
		\verb| |\\
		\verb| |\\
		\verb| |\\
		\end{tabularx}
				
\end{table}

\begin{table}[h!]
	\centering
		\begin{tabularx}{\columnwidth}{|X|}
			\hline
			\rowcolor{gray} \textbf{Q8.} \textit{Return all students with email address that are members of any department of a given university}\\
			\hline
			\scriptsize{\verb|SELECT ?X ?Y ?Z|}\\
			\scriptsize{\verb|WHERE | \{}\\
			\scriptsize{~~\verb|?X rdf:type ub:Student.|}\\
			\scriptsize{~~\verb|?Y rdf:type ub:Department.|}\\
			\scriptsize{~~\verb|?X ub:memberOf ?Y.|}\\
			\scriptsize{~~\verb|?Y ub:subOrganizationOf <http://www.University0.edu>.|}\\
			\scriptsize{~~\verb|?X ub:emailAddress ?Z|}\\
			\scriptsize{\}}\\
			\includegraphics[scale=0.67]{evaluation/lubm_q8}\\
			\hline
		\end{tabularx}	
	
		\begin{tabularx}{\columnwidth}{X}
		\verb| |\\
		\end{tabularx}	
		
		\begin{tabularx}{\columnwidth}{|X|}
			\hline
			\rowcolor{gray} \textbf{Q11.} \textit{Return all research groups of a given university}\\
			\hline
			\scriptsize{\verb|SELECT ?X|}\\
			\scriptsize{\verb|WHERE | \{}\\
			\scriptsize{~~\verb|?X rdf:type ub:ResearchGroup.|}\\
			\scriptsize{~~\verb|?X ub:subOrganizationOf <http://www.University0.edu>|}\\
			\scriptsize{\}}\\
			\includegraphics[scale=0.67]{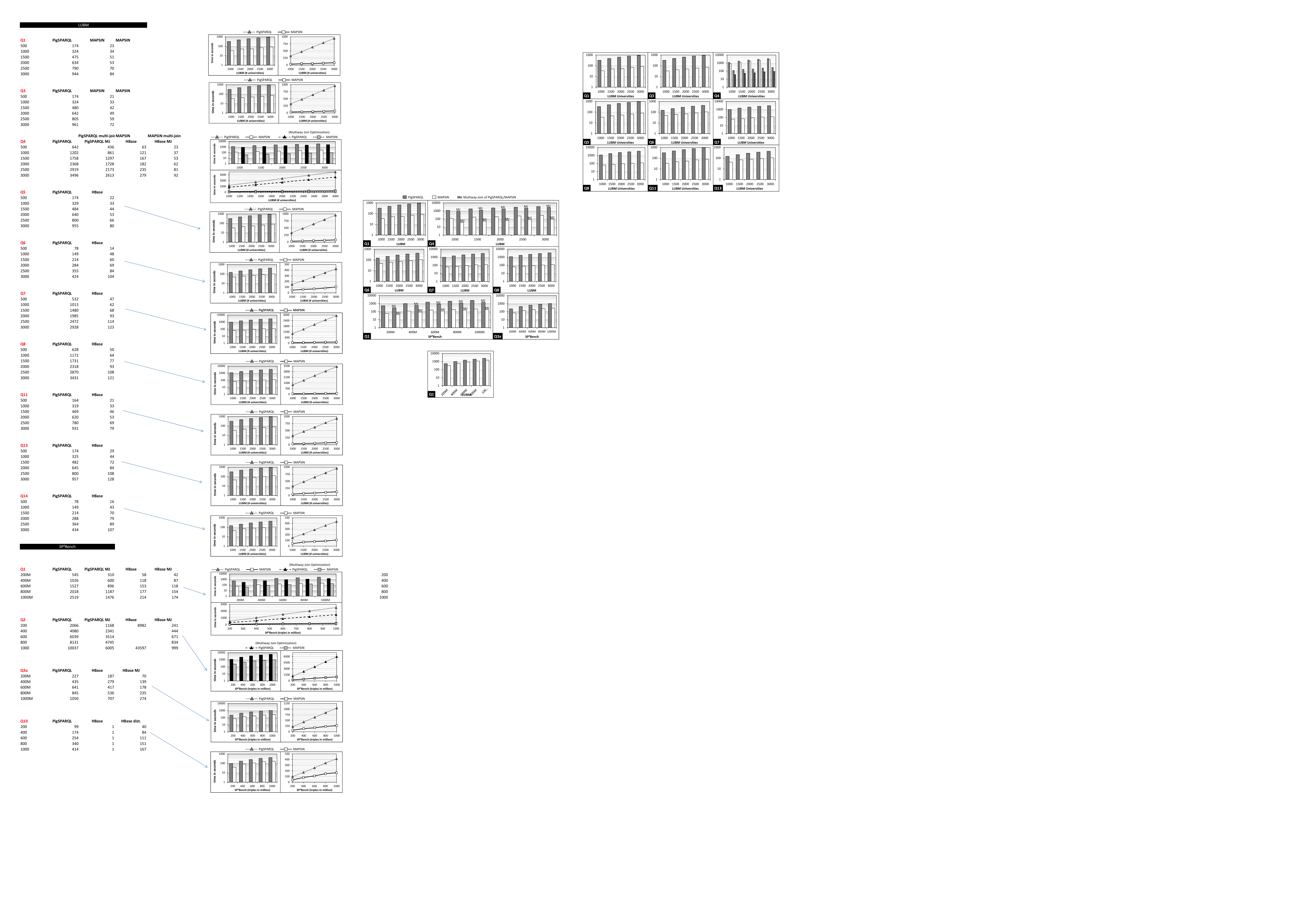}\\
			\hline
		\end{tabularx}
		
		\begin{tabularx}{\columnwidth}{X}
		\verb| |\\
		\verb| |\\
		\verb| |\\
		\verb| |\\
		\verb| |\\
		\verb| |\\
		\verb| |\\
		\verb| |\\
		\verb| |\\
		\verb| |\\
		\verb| |\\
		\verb| |\\
		\verb| |\\
		\verb| |\\
		\verb| |\\
		\verb| |\\
		\verb| |\\
		\verb| |\\
		\verb| |\\
		\verb| |\\
		\end{tabularx}
			
\newpage	
\end{table}		
		
\begin{table}[h!]
	\centering
	\begin{tabularx}{\columnwidth}{|X|}
			\hline
			\rowcolor{gray} \textbf{Q13.} \textit{Return all persons that are alumni of a given university}\\
			\hline
			\scriptsize{\verb|SELECT ?X|}\\
			\scriptsize{\verb|WHERE | \{}\\
			\scriptsize{~~\verb|?X rdf:type ub:Person.|}\\
			\scriptsize{~~\verb|<http://www.University0.edu> ub:hasAlumnus ?X|}\\
			\scriptsize{\}}\\
			\includegraphics[scale=0.67]{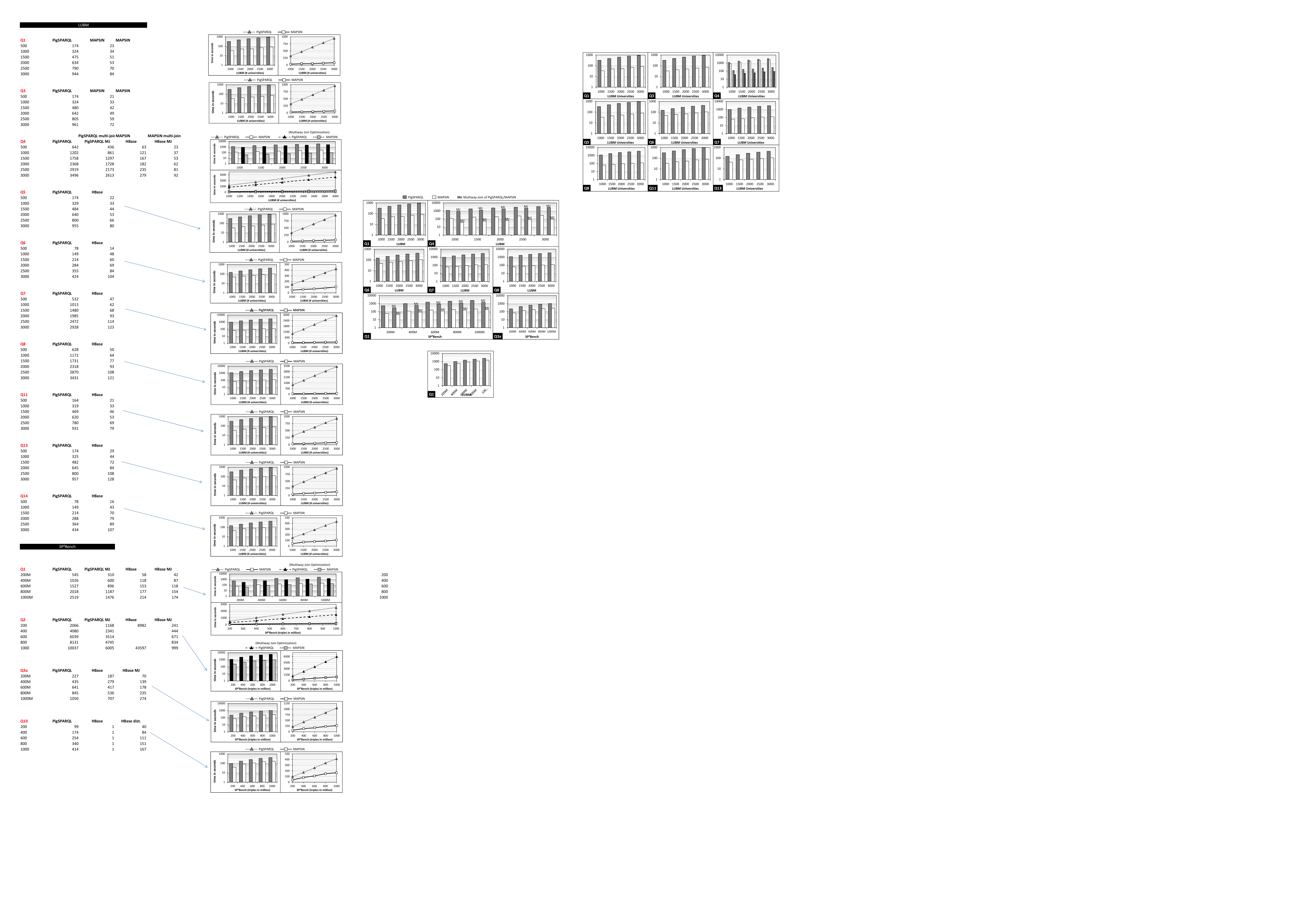}\\
			\hline
		\end{tabularx}
		
		\begin{tabularx}{\columnwidth}{X}
		\verb| |\\
		\end{tabularx}	
		
	\begin{tabularx}{\columnwidth}{|X|}
			\hline
			\rowcolor{gray} \textbf{Q14.} \textit{Return all undergraduate students}\\
			\hline
			\scriptsize{\verb|SELECT ?X|}\\
			\scriptsize{\verb|WHERE | \{}\\
			\scriptsize{~~\verb|?X rdf:type ub:UndergraduateStudent|}\\
			\scriptsize{\}}\\
			\includegraphics[scale=0.67]{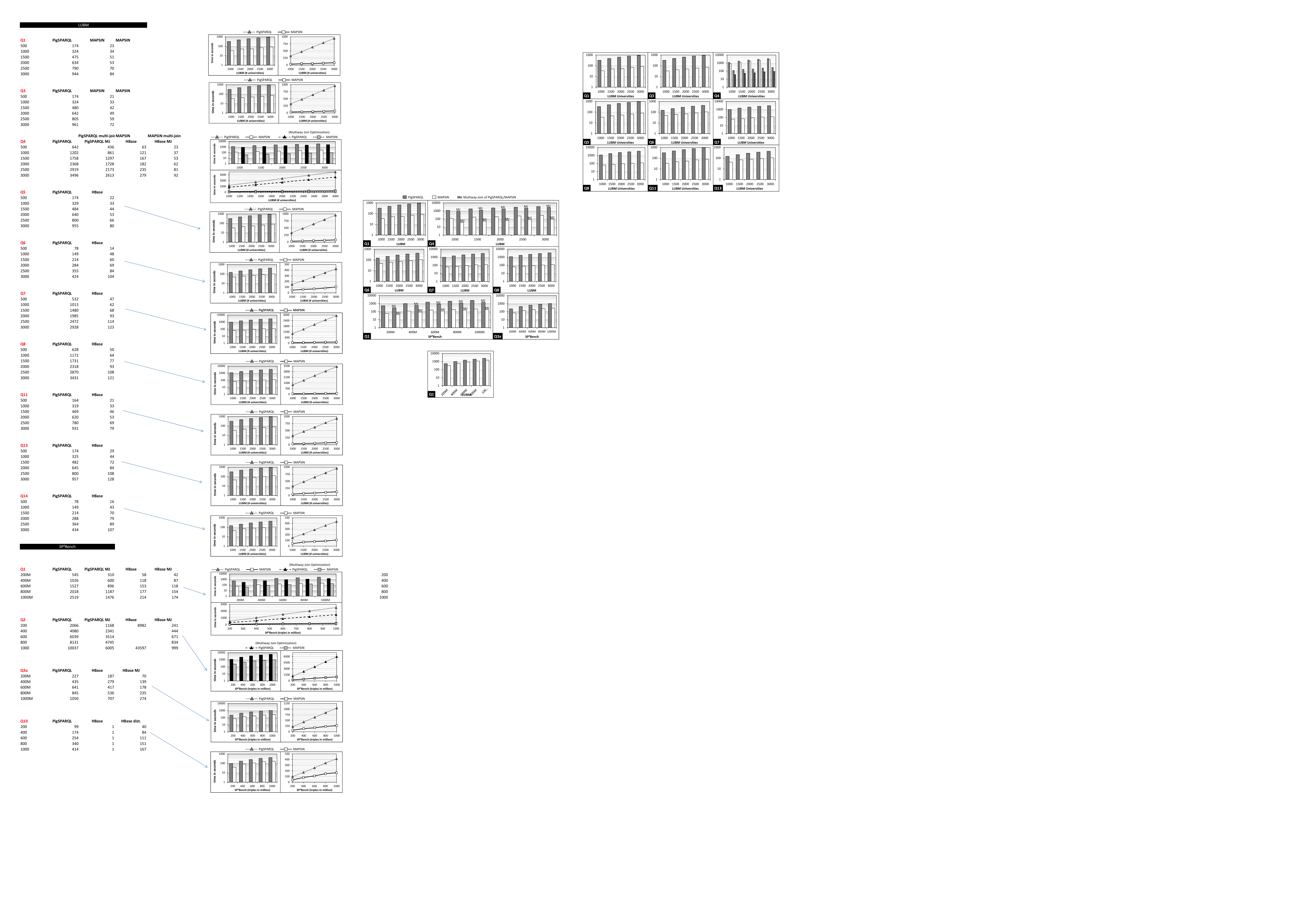}\\
			\hline
		\end{tabularx}

		\begin{tabularx}{\columnwidth}{X}
		\verb| |\\
		\verb| |\\
		\verb| |\\
		\verb| |\\
		\verb| |\\
		\verb| |\\
		\verb| |\\
		\verb| |\\
		\verb| |\\
		\verb| |\\
		\verb| |\\
		\verb| |\\
		\verb| |\\
		\verb| |\\
		\verb| |\\
		\verb| |\\
		\verb| |\\
		\verb| |\\
		\verb| |\\
		\verb| |\\
		\verb| |\\
		\verb| |\\
		\verb| |\\
		\verb| |\\
		\end{tabularx}
\end{table}

\clearpage
\section{SP$^2$Bench}
\label{appendix-sp2bench}
\begin{table}[h!]
	\centering
		\begin{tabularx}{\columnwidth}{|X|}
			\hline
			\rowcolor{gray} \textbf{Q1.} \textit{Return the year of publication of ''Journal 1 (1940)''}\\
			\hline
			\scriptsize{\verb|SELECT ?yr|}\\
			\scriptsize{\verb|WHERE | \{}\\
			\scriptsize{~~\verb|?journal rdf:type bench:Journal.|}\\
			\scriptsize{~~\verb|?journal dc:title "Journal 1 (1940)"|\textasciicircum \textasciicircum \verb|xsd:string.|}\\
			\scriptsize{~~\verb|?journal dcterms:issued ?yr|}\\
			\scriptsize{\}}\\
			\includegraphics[scale=0.67]{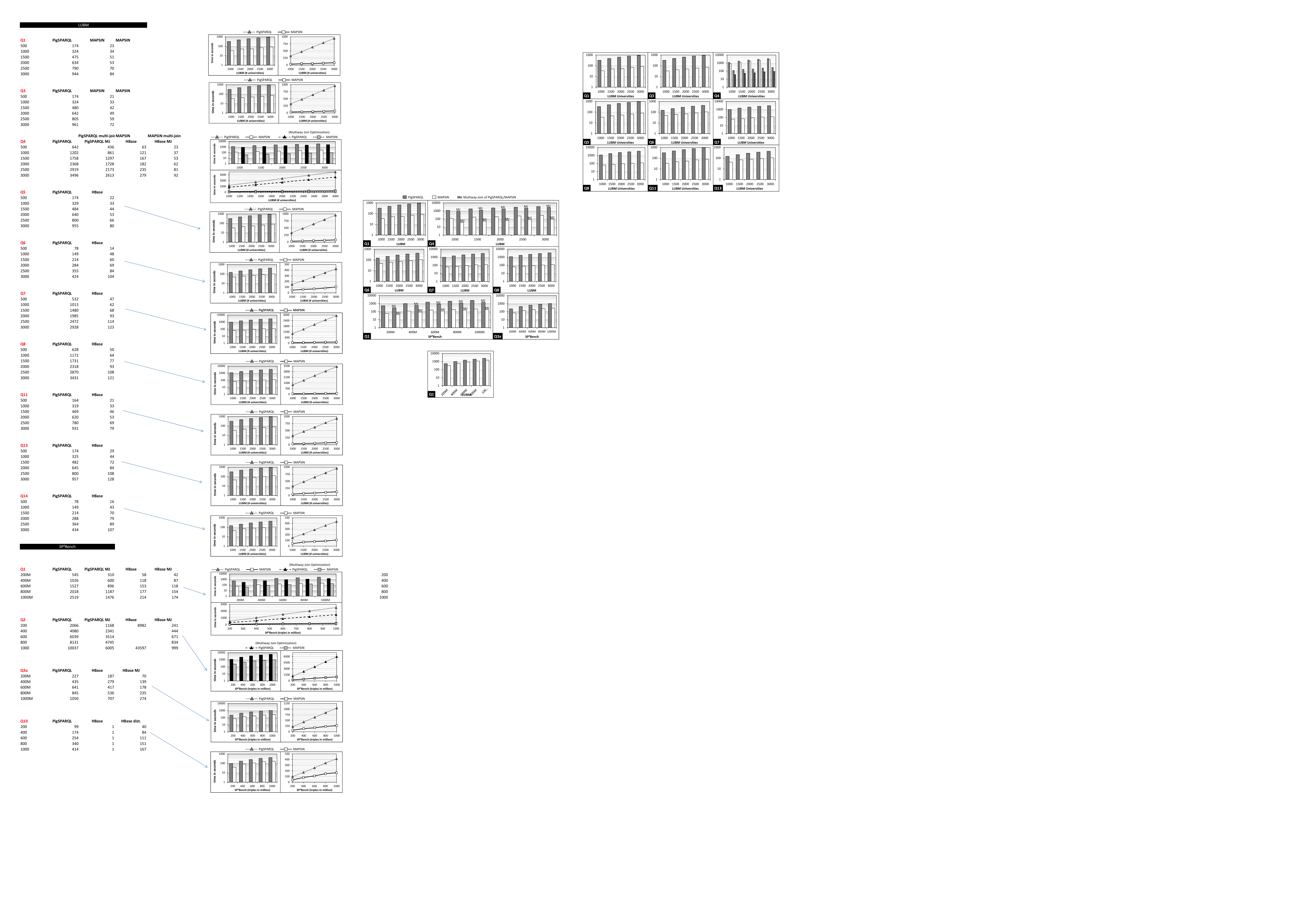}\\
			\hline	
		\end{tabularx}
		
		\begin{tabularx}{\columnwidth}{X}
		\verb| |\\
		\end{tabularx}
		
		\begin{tabularx}{\columnwidth}{|X|}
			\hline
			\rowcolor{gray} \textbf{Q2.} \textit{Return all inproceedings with the given properties (slightly modified version of the original query without OPTIONAL and ORDER BY)}\\
			\hline
			\scriptsize{\verb|SELECT ?inproc ?author ?booktitle ?title|}\\
			\scriptsize{~~~~~~~~~~\verb|?proc ?ee ?page ?url ?yr|}\\
			\scriptsize{\verb|WHERE | \{}\\
			\scriptsize{~~\verb|?inproc rdf:type bench:Inproceedings.|}\\
			\scriptsize{~~\verb|?inproc dc:creator ?author.|}\\
			\scriptsize{~~\verb|?inproc bench:booktitle ?booktitle.|}\\
			\scriptsize{~~\verb|?inproc dc:title ?title.|}\\
			\scriptsize{~~\verb|?inproc dcterms:partOf ?proc.|}\\
			\scriptsize{~~\verb|?inproc rdfs:seeAlso ?ee.|}\\
			\scriptsize{~~\verb|?inproc swrc:pages ?page.|}\\
			\scriptsize{~~\verb|?inproc foaf:homepage ?url.|}\\
			\scriptsize{~~\verb|?inproc dcterms:issued ?yr|}\\
			\scriptsize{\}}\\
			\includegraphics[scale=0.67]{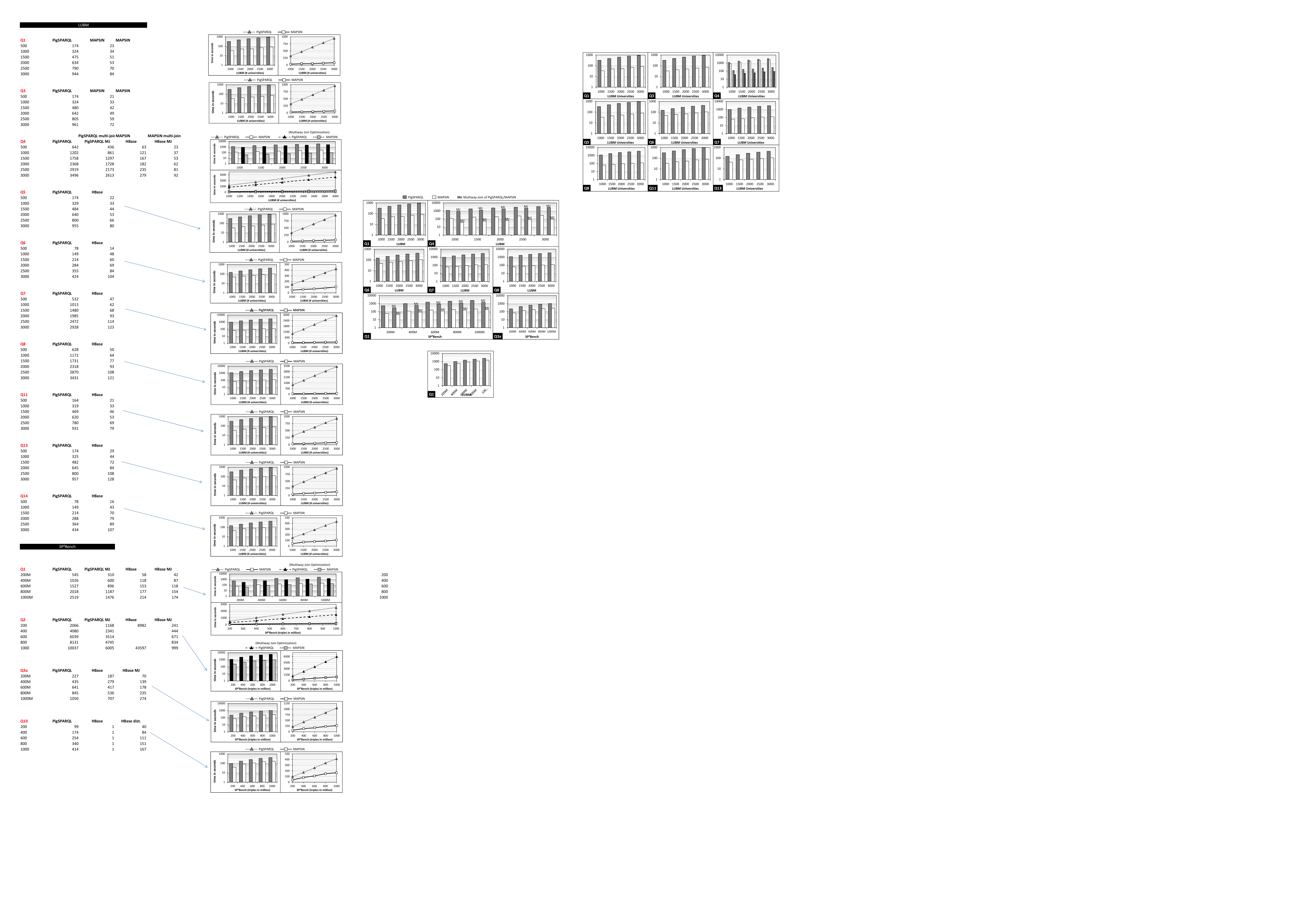}\\
			\hline	
		\end{tabularx}
		
		\begin{tabularx}{\columnwidth}{X}
		\verb| |\\
		\verb| |\\
		\verb| |\\
		\verb| |\\
		\verb| |\\
		\verb| |\\
		\verb| |\\
		\verb| |\\
		\verb| |\\
		\verb| |\\
		\verb| |\\
		\verb| |\\
		\verb| |\\
		\end{tabularx}
		
\end{table}

\begin{table}[htbp]
	\centering
	
		\begin{tabularx}{\columnwidth}{|X|}
			\hline
			\rowcolor{gray} \textbf{Q3a.} \textit{Return all articles with property swrc:pages}\\
			\hline
			\scriptsize{\verb|SELECT ?article|}\\
			\scriptsize{\verb|WHERE | \{}\\
			\scriptsize{~~\verb|?article rdf:type bench:Article.|}\\
			\scriptsize{~~\verb|?article ?property ?value|}\\
			\scriptsize{~~\verb|FILTER (?property=swrc:pages)|}\\
			\scriptsize{\}}\\
			\includegraphics[scale=0.67]{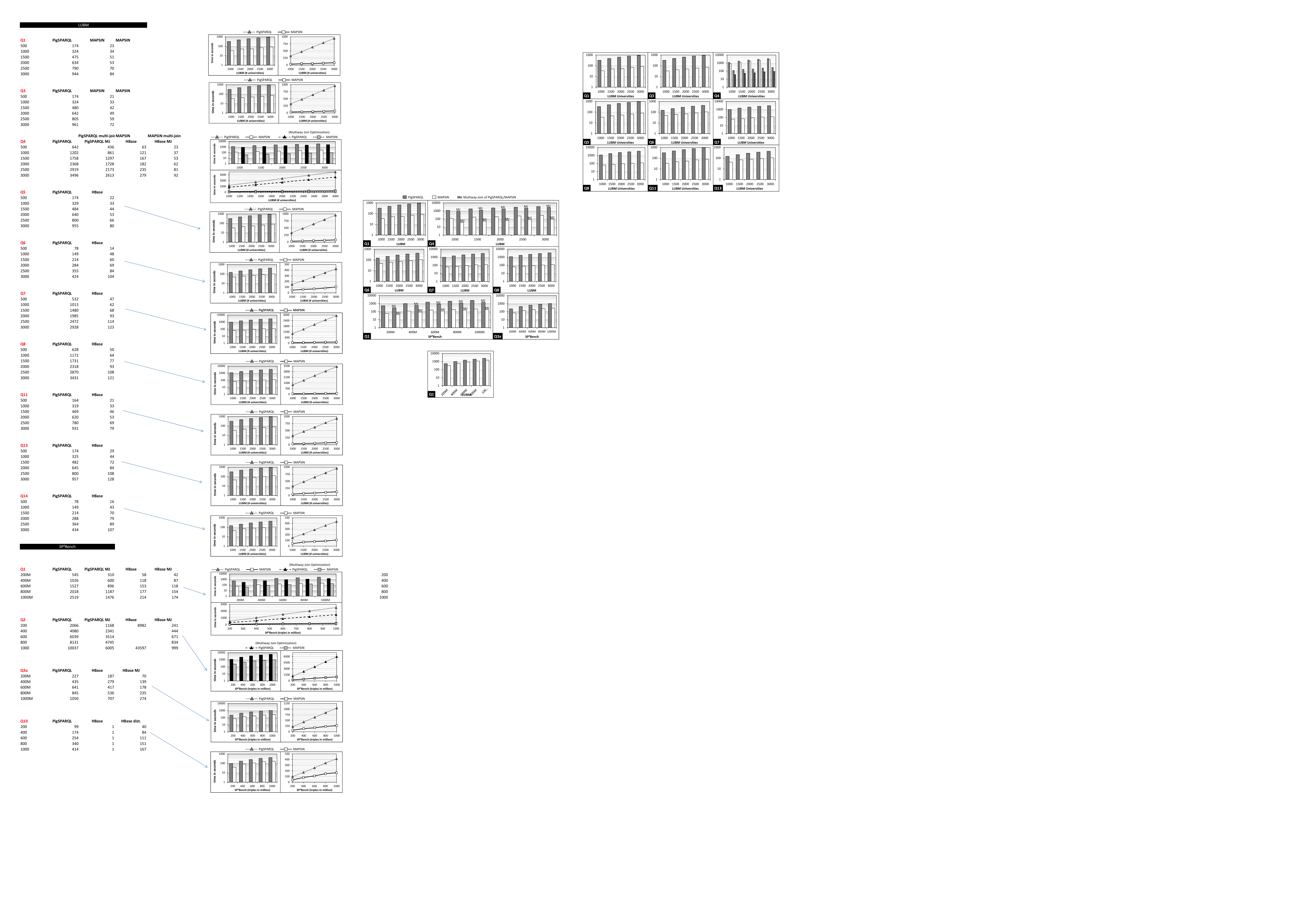}\\
			\hline	
		\end{tabularx}	
	
		\begin{tabularx}{\columnwidth}{X}
		\verb| |\\
		\end{tabularx}	
	
		\begin{tabularx}{\columnwidth}{|X|}
			\hline
			\rowcolor{gray} \textbf{Q10.} \textit{Return all subjects that stand in any relation to person ''Paul Erd\"os''}\\
			\hline
			\scriptsize{\verb|SELECT ?subj ?pred|}\\
			\scriptsize{\verb|WHERE | \{}\\
			\scriptsize{~~\verb|?subj ?pred person:Paul_Erdoes|}\\
			\scriptsize{\}}\\
			\includegraphics[scale=0.67]{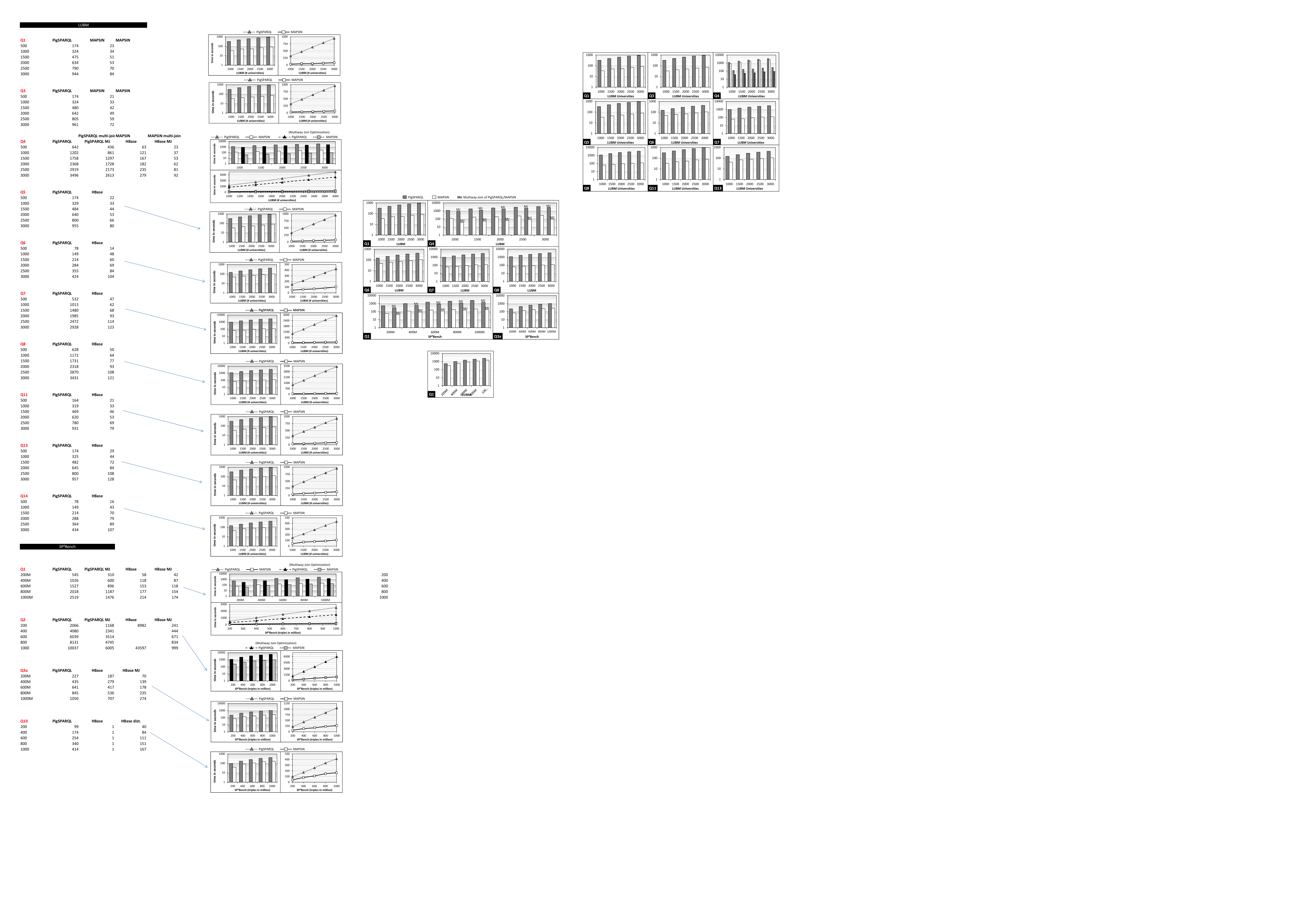}\\
			\hline
		\end{tabularx}
		
		\begin{tabularx}{\columnwidth}{X}
		\verb| |\\
		\verb| |\\
		\verb| |\\
		\verb| |\\
		\verb| |\\
		\verb| |\\
		\verb| |\\
		\verb| |\\
		\verb| |\\
		\verb| |\\
		\verb| |\\
		\verb| |\\
		\verb| |\\
		\verb| |\\
		\verb| |\\
		\verb| |\\
		\verb| |\\
		\verb| |\\
		\verb| |\\
		\verb| |\\
		\verb| |\\
		\verb| |\\
		\verb| |\\
		\verb| |\\
		\verb| |\\
		\verb| |\\
		\verb| |\\
		\verb| |\\
		\end{tabularx}
		
\end{table}


\end{document}